\documentclass[9pt,twocolumn,a4paper]{extarticle}

\usepackage[a4paper,
            textwidth=42pc,
            textheight=636pt,      %
            columnsep=2pc,
            headheight=11pt,
            headsep=11pt,
            footskip=22pt,
            centering]{geometry}

\usepackage[T1]{fontenc}
\usepackage[utf8]{inputenc}
\usepackage{newtxtext}
\usepackage{newtxmath}
\usepackage{microtype}

\usepackage{amsmath}          %
\usepackage{graphicx}
\usepackage{booktabs}
\usepackage[font=small,labelfont=bf]{caption}
\usepackage{subcaption}
\usepackage{xcolor}

\usepackage{titlesec}
\titleformat{\section}{\normalfont\large\bfseries}{\thesection.}{0.5em}{}
\titleformat{\subsection}{\normalfont\normalsize\bfseries}{\thesubsection.}{0.5em}{}
\titlespacing*{\section}{0pt}{1.6ex plus .5ex minus .2ex}{0.9ex plus .2ex}
\titlespacing*{\subsection}{0pt}{1.2ex plus .4ex minus .2ex}{0.6ex plus .2ex}
\titlespacing*{\paragraph}{0pt}{1.0ex plus .3ex minus .2ex}{0.6em}

\usepackage[numbers,sort&compress]{natbib}

\usepackage[colorlinks=true,
            linkcolor={blue!60!black},
            citecolor={blue!60!black},
            urlcolor={blue!60!black}]{hyperref}

\graphicspath{{./}}

\begin{document}
\setcounter{secnumdepth}{2}

\twocolumn[{%
\begin{center}
    {\LARGE\bfseries StippleDiffusion: Capacity-Constrained Stippling using Controlled Diffusion\par}
    \vspace{1.2em}
    {\large Ofir Gilad$^{1}$ \quad Aleksander Plocharski$^{2}$ \quad Przemyslaw Musialski$^{3}$ \quad Andrei Sharf$^{1}$\par}
    \vspace{0.6em}
    {\small $^1$Ben-Gurion University of the Negev, Israel \quad
            $^2$Warsaw University of Technology, Poland \quad
            $^3$New Jersey Institute of Technology, USA\par}
    \vspace{1.2em}
    \includegraphics[width=0.54\textwidth]{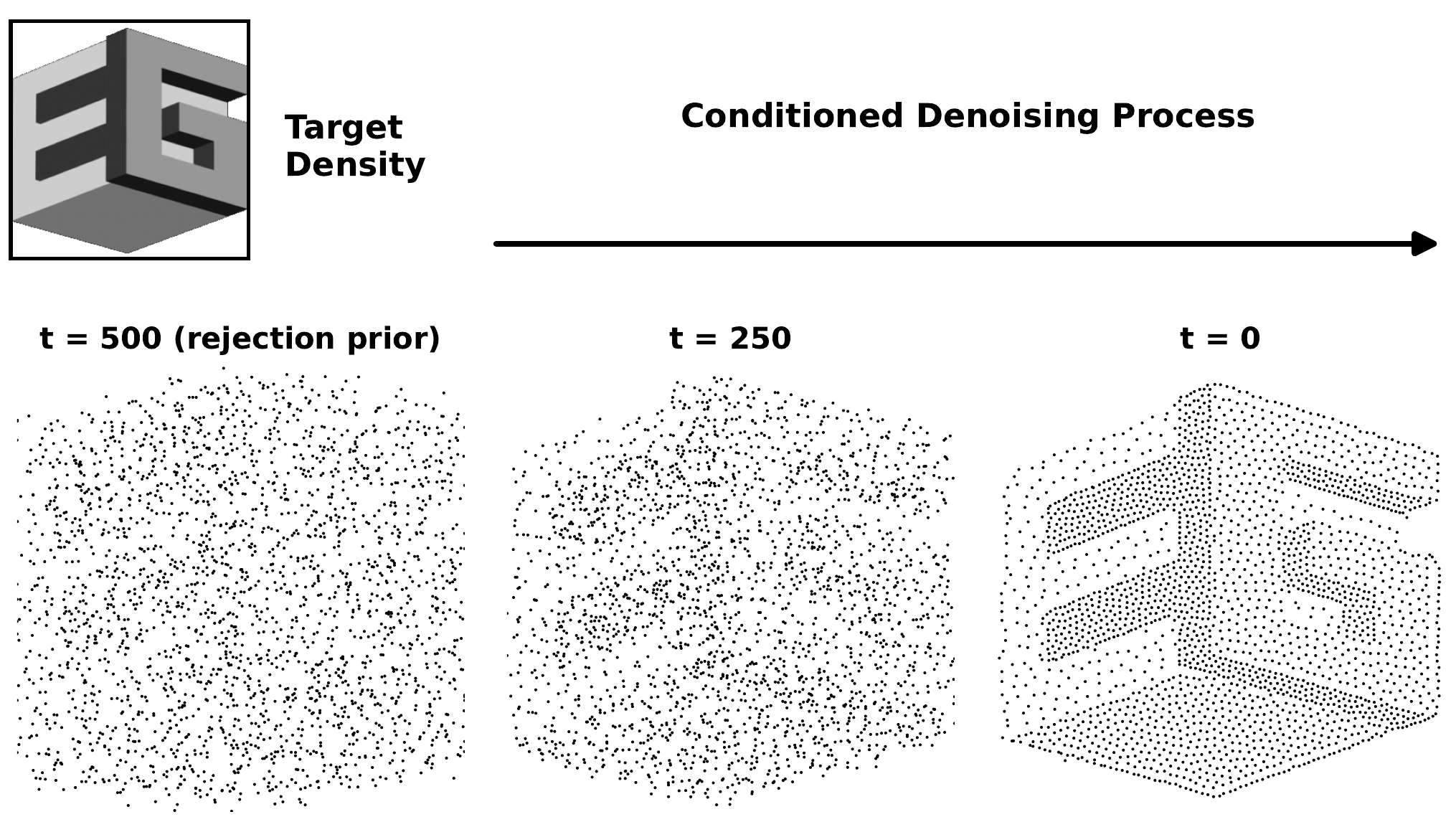}\hfill
    \includegraphics[width=0.44\textwidth]{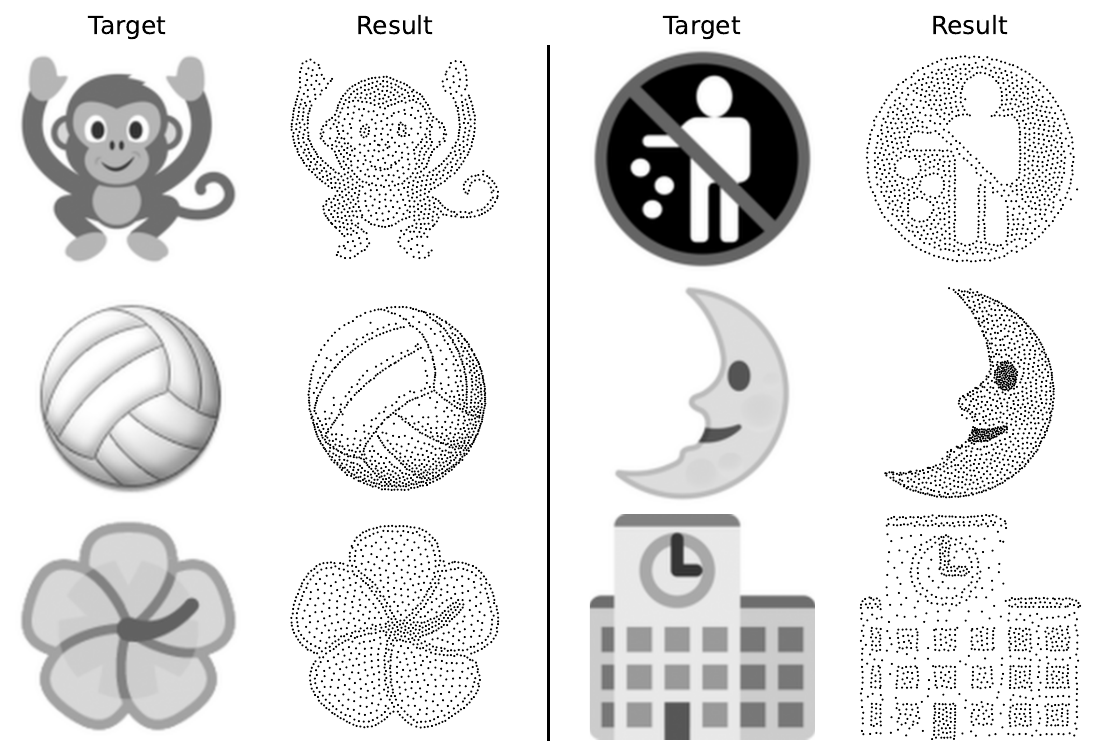}
    \captionof{figure}{
    StippleDiffusion generates capacity-constrained stipple patterns by conditioning a diffusion model on a target image. Left: controlled denoising progressively refines the point distribution while preserving local point repulsion and density. Right: results for diverse target images, with generated stipple patterns accurately matching the target shapes, point capacity, and distributions.
    }
    \label{fig:teaser}
\end{center}
\vspace{0.4em}
\hrule
\vspace{0.8em}
{\small
\noindent\textbf{Abstract}\par\vspace{0.3em}
\noindent
Stipple patterns are point sets whose local density follows a target image. They are traditionally generated by iterative optimization, which must be re-run for every target and provides no gradients. Learned point-diffusion methods, in contrast, generate points unconditionally. A model can be trained to reproduce one density, but it takes no input and cannot adapt to a new one.
We present a conditional point-diffusion sampler that takes a target image and generates a stipple pattern in the style of the sampler that produced its training exemplars. The model operates on an optimal-transport grid and adds a jointly trained conditioning branch with per-point image features. At inference, a density-aware initialization is refined by truncated denoising. We evaluate it on four datasets, with Weighted Voronoi Stippling and Gaussian Blue Noise as exemplar samplers. Each learned variant reproduces the characteristic strengths of its exemplar sampler across the metrics. It improves on that sampler in HOT energy, centroid displacement and EMD, and falls short of it in CVT energy and spatial measure. Its inference cost is set by a fixed number of denoising steps and grows little with the point budget. Our sampler is differentiable, enabling downstream optimization of the requested density to correct dot gain in rendered stipples.
\par\vspace{0.6em}
\par}
\vspace{0.8em}
\hrule
\vspace{1.6em}
}]

\section{Introduction}

Generating high-quality 2D point sets with prescribed spatial properties is a fundamental problem in computer graphics, with applications including rendering, object placement, and image stippling. Classical approaches typically rely on iterative optimization or specialized sampling strategies to produce structured distributions such as blue noise. While effective, these methods can be computationally expensive and difficult to adapt to multiple constraints. In particular, a new target density generally requires a separate optimization process, limiting their scalability and suitability for interactive or large-scale generation.

Recently, denoising diffusion probabilistic models have emerged as a powerful framework for learning complex geometric distributions directly from data~\cite{10.1145/3610548.3618243}. However, applying diffusion models to scattered point sets remains challenging because standard convolutional architectures are designed for grid-structured data rather than unordered points. To address this issue, \citet{10.1145/3610548.3618243}, building on the linear optimal-transport assignment of \citet{10.1145/2070781.2024192}, represent point sets on an OT grid by assigning each point to a grid cell and encoding its position as a local offset. This representation enables efficient convolutional processing, but its grid-based formulation can make it challenging to capture fine-scale spatial relationships and structured point distributions.

Conditioning diffusion models on spatially varying targets introduces an additional challenge. Although standard conditioning mechanisms such as ControlNet~\cite{zhang2023adding} can be applied to OT-grid representations, enforcing fine-grained density and geometric constraints during generation remains difficult. The model must maintain globally well-distributed points while simultaneously capturing high-frequency variations in target density and shape. Poorly balanced conditioning can lead to outliers, uneven densities, residual clustering, and loss of fine geometric structure.

In this paper, we learn the stippling process of a chosen sampler. Given exemplars from Weighted Voronoi Stippling (WVS) or Gaussian Blue Noise (GBN), our model stipples unseen images in that sampler's style, for different point budgets. It starts from a density-weighted initialization and refines it by truncated denoising, inspired by SDEdit~\cite{meng2022sdeditguidedimagesynthesis} (Figure~\ref{fig:teaser}).

Our work makes the following contributions:
\begin{itemize}

\item We propose a conditional, example-based stippler. For each dataset and exemplar sampler, we train one network on images paired with their exemplar stipples at a single, fixed point budget. At inference, it generates stipples for unseen images, including at point budgets other than the training one, without per-target optimization. The learned variants reproduce the characteristic strengths of their exemplar samplers across the metrics.

\item We combine an optimal-transport grid with a jointly trained conditioning branch (per-point image features, gated injection). In the ablation, this lowers capacity deviation from 0.063 to 0.026 compared with a standard ControlNet on the frozen base model. A density-aware initialization with truncated denoising halves the inference time at comparable quality.

\item Our sampler is differentiable from the generated stipple back to the requested density, enabling downstream optimization. We demonstrate this for dot-gain correction by back-propagating through a single denoising step.

\end{itemize}

\begin{figure*}[t]
    \centering
    \includegraphics[width=0.9\linewidth]{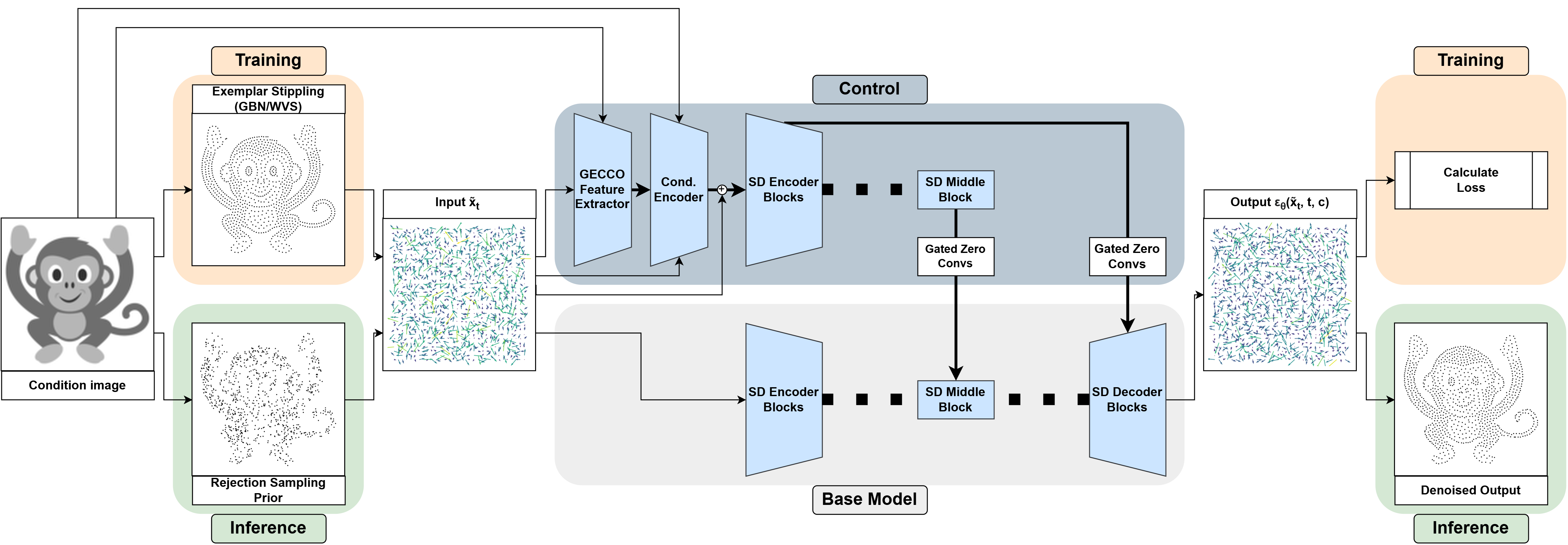}
    \caption{Pipeline. Given a condition image, training (orange) uses an exemplar stipple generated by GBN or WVS, while inference (green) initializes the point set using density-weighted rejection sampling. The point set is converted to OT offsets and noised before entering the diffusion network, which is conditioned on image features. During training, the predicted noise is compared with the known noise to compute the loss; at inference, it is used for 500 denoising steps to generate the final stipple.
    }
    \label{fig:pipeline}
\end{figure*}

\section{Related Work}
\label{sec:related}

\textbf{Blue Noise Sampling.}
Blue-noise distributions are characterized by a power spectrum with suppressed low frequencies and energy concentrated in an isotropic high-frequency annulus. They arise in biologically optimized arrangements such as the retinal cone mosaic \cite{Yellott1982}, and are widely used in digital halftoning \cite{Ulichney1987}, image stippling \cite{Deussen2000}, and Monte Carlo integration \cite{SubrKautz2013, Pilleboue2015}. Existing construction methods include kernel-based formulations \cite{10.1145/2010324.1964943}, pair-correlation matching \cite{Oztireli2012}, optimal transport \cite{deGoes:2012:BNOT, Qin2017, Paulin2020}, fast approximations \cite{Nader2018}, and Gaussian variants \cite{Ahmed2022Gaussian}. Tile-based methods instead precompute reusable patches, trading computation for memory \cite{Ostromoukhov2004, Kopf2006, Ostromoukhov2007, Wachtel2014, Ahmed2016LDBN}.

These methods use different mechanisms to enforce point separation, including Lloyd iterations \cite{article1992, Deussen2000, 10.1145/508530.508537, 10.1145/1576246.1531392, 10.1145/1730804.1730985, XU2011510, 6197186}, electrostatic forces \cite{https://doi.org/10.1111/j.1467-8659.2010.01716.x}, farthest-point optimization \cite{Schlomer2011Accur-17722}, and sampling on regular lattices \cite{10.1145/3072959.3073588}. While these approaches produce high-quality distributions, many require iterative computation whose cost increases with the number of points. We refer to \citet{Yan2015} and \citet{https://doi.org/10.1111/cgf.13653} for broader surveys of blue-noise sampling and sample correlations.

Blue-noise sampling produces well-distributed points at approximately uniform density, while stippling extends this idea to spatially varying densities by modulating point density according to an input image. This concept has also been extended to color stippling by formulating it as a multi-class blue-noise sampling problem \cite{10.1145/3306214.3338606}.

\textbf{Image-Conditioned Diffusion Models for Stippling.}
Few works have addressed stippling with neural networks. The closest is the feed-forward stippling network of \citet{10.1007/978-3-030-89029-2_24}, which predicts color stipple positions at a fixed output size using a spectrum-based loss. In contrast, our method takes only the target image as input, supports different point budgets through the size of its OT grid, and preserves the prescribed point mass per region. We therefore consider this work related but not directly comparable to ours.

Learned point samplers with fast inference and point-count generalization have also been explored. Deep Point Correlation Design~\cite{10.1145/3355089.3356562} optimizes a composition of distance-based filters under a user-specified correlation loss and generalizes across point counts. However, its conditioning is based on spatial or spectral statistics rather than an image-defined density field, and it does not address capacity-constrained stippling.

Diffusion models have recently enabled powerful conditional generation. SDEdit~\cite{meng2022sdeditguidedimagesynthesis} uses a structural prior and a truncated reverse process for image-to-image refinement, while ControlNet~\cite{zhang2023adding} introduces spatial conditioning through parallel encoding branches. Although developed for image grids, these ideas provide useful principles for conditional geometric generation.

Applying diffusion to scattered point sets presents an architectural challenge. \citet{10.1145/3610548.3618243} address this by matching points to a regular grid through optimal transport, enabling standard convolutional architectures for point diffusion. While effective for unconditional point generation, extending this representation to high-frequency, spatial conditioning remains challenging, as strong spatial guidance can disrupt the underlying point distribution.

GECCO~\cite{tyszkiewicz2023geccogeometricallyconditionedpointdiffusion} instead uses coordinate-based features for geometrically conditioned point diffusion, while gated architectures such as highway networks~\cite{srivastava2015highwaynetworks} provide a mechanism for integrating auxiliary features while preserving the underlying representation.

Our method builds on these ideas to enable high-frequency spatial conditioning for image stippling.

\section{Method}
\label{sec:method}

\paragraph{Architecture} Our method builds upon diffusion models for 2D point-set generation, which progressively transform noisy point representations into structured distributions through iterative denoising. We use the optimal-transport (OT) point-set diffusion model of \citet{10.1145/3610548.3618243}, which represents unordered point sets on a grid to enable convolutional processing (see Figure~\ref{fig:pipeline}).

The denoising process consists of a sequence of steps indexed by $t$. At each step, a denoising network $\epsilon_{\theta}$ predicts the noise in the noisy grid $\tilde{x}_t$, conditioned on the embedded timestep $t$.
The network is a U-Net~\cite{ronneberger2015unetconvolutionalnetworksbiomedical}, with two convolutional residual blocks at each level and factor-of-two downsampling between consecutive levels.

To enable spatial conditioning, our architecture introduces a conditioning branch in parallel with the base U-Net, inspired by ControlNet~\cite{zhang2023adding}. Unlike the original ControlNet paradigm, which keeps the pretrained base model frozen, we randomly initialize both the base U-Net and the conditioning branch and train them jointly from scratch.

Rather than using standard Zero Convolutions to inject conditioning features, we replace them with Gated Zero Convolutions~\cite{srivastava2015highwaynetworks}. The motivation is to prevent strong spatial conditioning from disrupting the geometric regularity of the point distribution, particularly during early denoising. At the middle and decoding layers of the U-Net, a sigmoid-gated $1\times1$ projection controls the strength of the injected features, allowing the network to gradually and selectively incorporate spatial guidance while preserving the underlying point-distribution structure.

The conditioning input $\mathbf{c}$ is rebuilt at every denoising step from the current noisy offsets (Figure~\ref{fig:pipeline}). Following GECCO~\cite{tyszkiewicz2023geccogeometricallyconditionedpointdiffusion}, a lightweight CNN processes the full-resolution input image into a dense feature map, and $16$ feature channels are read from this map at the continuous positions of the current noisy points by bilinear interpolation. These are concatenated with the noisy offsets themselves (2 channels) and the downsampled density map (1 channel), giving a 19-channel condition on the $N\times N$ grid. A convolutional Conditioning Encoder with SiLU activations and no downsampling lifts it from 19 to 32, 64, and 128 channels, with dilation rates up to $4$ to capture broader spatial context while preserving spatial resolution.

The combined network learns a time-dependent noise model $\epsilon_{\theta}(\tilde{x}_t, t, \mathbf{c})$, where $\tilde{x}_t$ is obtained by adding noise $\epsilon_t$ to the offset grid $x_0$ of an exemplar point set:
$$
\tilde{x}_t = \sqrt{\bar{\alpha}_t}\,x_0 + \sqrt{1-\bar{\alpha}_t}\,\epsilon_t,
$$
at each time step $t$, conditioned on our spatial features $\mathbf{c}$. Here $\epsilon_t\sim\mathcal{N}(0,\mathbf{I})$ is Gaussian noise, and $\bar{\alpha}_t=\prod_{\tau=1}^{t}(1-\beta_\tau)$ is the cumulative product of a linear noise schedule whose $\beta_\tau$ rise from $10^{-4}$ to $10^{-2}$ over $T=1000$ steps. It sets how much of the clean offsets remains in $\tilde{x}_t$, falling from nearly $1$ at $t=0$ to nearly $0$ at $t=T$, where $\tilde{x}_t$ is almost pure noise.

At inference, we initialize the reverse process from a partially noised, rejection-sampled point set using a truncated SDEdit schedule (see Inference below) to reduce runtime. Training always uses the full diffusion schedule, and the same model can also sample from pure noise.

\paragraph{Optimal-transport grid representation}
Our network operates on a regular 2D grid rather than directly on the raw point set, following the representation of \citet{10.1145/3610548.3618243}. For a budget of $n$ points, we partition the unit square $[0,1]^2$ into an $N\times N$ grid with $N=\sqrt{n}$ (e.g., $N=32$ for $1024$ points), with cell centers $g_{u,v}$. Each cell is assigned exactly one point $s_{u,v}$ through a one-to-one linear optimal-transport matching~\cite{10.1145/2070781.2024192} that minimizes the total squared distance between cell centers and assigned points. We store the continuous offset
$$
o_{u,v}=s_{u,v}-g_{u,v}
$$
in each cell, yielding a two-channel $N\times N$ representation $x_0$. Standard 2D convolutions can then process the point set as an image, with neighboring cells providing an approximate local neighborhood of nearby samples. Importantly, the representation does not quantize the points: offsets remain continuous, samples retain their coordinates in $[0,1]^2$, and generated points are never snapped to grid locations.

The grid depends only on the point budget, not on the input image. The grayscale image defines the target density, with darker regions corresponding to higher density, and is mapped to the same unit square regardless of its resolution or aspect ratio. Each axis is normalized independently, allowing non-square images to be processed at native resolution without padding. The same architecture supports square $512\times512$ images and non-square images with a longer side of up to $1024$ pixels. The image enters the network in two forms: a downsampled $N\times N$ density map aligned with the OT offsets, and a full-resolution image from which features are sampled at the continuous point coordinates.

\paragraph{Training}
For each batch of size $B$, we pair each training image with an exemplar point set generated offline by a stippling sampler, whose point distribution the network learns to reproduce. Thus, the choice of exemplar sampler determines the point-distribution characteristics learned by the model, while the input image determines where points are placed.

We train separate models using exemplars generated by Gaussian Blue Noise (GBN)~\cite{Ahmed2022Gaussian} and Weighted Voronoi Stippling (WVS)~\cite{10.1145/508530.508537}. We chose these samplers because they provide a good balance between generation efficiency and stipple quality, allowing us to generate training exemplars at practical computational cost while maintaining well-structured point distributions. The framework is not tied to these samplers and can similarly learn distributions produced by other methods, provided that their optimization converges reliably and can generate exemplars efficiently at the scale required for training.

The BNOT sampler~\cite{deGoes:2012:BNOT} also produces capacity-constrained stipples using an optimal-transport formulation. However, its relatively expensive iterative optimization makes it impractical for generating the large number of training exemplars required by our approach.

We use GBN as the default for ablations and runtime evaluation because it produces visually pleasing stipples while providing a challenging distribution for the learned sampler to reproduce.

Each model is trained at a fixed grid size determined by its dataset's point budget (e.g., $N=32$ for $1024$ points and $N=40$ for $1600$ points). For each training image (Figure~\ref{fig:pipeline}, orange), the exemplar points are converted once during data loading into an exemplar offset grid $x_0$ using the OT assignment described above. The image provides two conditioning signals: the full-resolution image, from which image features are extracted, and a density map downsampled to the $N\times N$ grid.

For each batch element $i$, we sample a timestep $t_i \sim \mathcal{U}(0,T)$ from the full training schedule ($T=1000$) and add Gaussian noise $\epsilon^{(i)}$ to its exemplar offset grid $x^{(i)}_0$, yielding the noisy representation $\tilde{x}^{(i)}_{t_i}$. The network then predicts the added noise conditioned on the corresponding spatial features.

We optimize a combination of Min-SNR-$\gamma$ weighted MSE~\cite{hang2024efficientdiffusiontrainingminsnr} and a Kernel Density Estimation (KDE) loss:
\begin{multline*}
\mathcal{L}(\theta) = \\
\frac{1}{B}\sum_{i=1}^{B}\Big[\, w(t_i)\,\big\|\epsilon_\theta(\tilde{x}^{(i)}_{t_i},t_i,\mathbf{c}^{(i)})-\epsilon^{(i)}\big\|^2 + \lambda_{\mathrm{KDE}}\,\mathcal{L}_{\mathrm{KDE}}(\hat{x}^{(i)}_{0},x^{(i)}_{0}) \Big].
\end{multline*}
Here, $i$ indexes the batch, $x^{(i)}_0$ denotes the exemplar offset grid, $\tilde{x}^{(i)}_{t_i}$ is its noisy version at timestep $t_i$ with noise $\epsilon^{(i)}$, and $\mathbf{c}^{(i)}$ denotes the corresponding spatial condition. The predicted clean grid $\hat{x}^{(i)}_0$ is obtained from the network's noise prediction.

The first term is the standard noise-prediction loss, weighted by
$w(t)=\min(\mathrm{SNR}(t),\gamma)/\mathrm{SNR}(t)$, where $\mathrm{SNR}(t)=\bar{\alpha}_{t}/(1-\bar{\alpha}_{t})$ and $\gamma=5.0$. This weighting balances the contribution of different diffusion timesteps~\cite{hang2024efficientdiffusiontrainingminsnr}.

The second term operates on the predicted point distribution rather than the noise. Specifically, the predicted noise provides a closed-form estimate of the clean offsets,
$$
\hat{x}^{(i)}_0=\big(\tilde{x}^{(i)}_{t_i}-\sqrt{1-\bar{\alpha}_{t_i}}\,\epsilon_\theta\big)/\sqrt{\bar{\alpha}_{t_i}},
$$
which is decoded into point positions, as is the exemplar $x^{(i)}_0$. For each point set, we construct a coarse density image on a fixed $32\times32$ grid, independent of the OT grid resolution. Each cell accumulates a Gaussian kernel with one-cell standard deviation centered at every point, and the resulting image is normalized to have unit mean. The KDE loss $\mathcal{L}_{\mathrm{KDE}}$ is the mean squared difference between the predicted and exemplar density images. It therefore penalizes local regions that contain too few or too many points, providing a direct density constraint that the per-point noise loss captures only indirectly.

We apply the KDE term only for $t_i<0.4T$, where the estimate $\hat{x}_0$ is sufficiently reliable, and set $\lambda_{\mathrm{KDE}}=0.8$, linearly ramped over the first five epochs.

\paragraph{Inference}
Given a novel input image, we first draw the $n$ points of the chosen point budget by density-weighted rejection sampling from the image (Figure~\ref{fig:pipeline}, green), making darker pixels proportionally more likely to receive points. This provides a structural prior that accelerates convergence compared with initialization from pure Gaussian noise.

The sampled points are converted to an offset grid on an $N\times N$ grid, with $N=\sqrt{n}$, using the same OT assignment as during training. Since $N$ need not match the training resolution, a single checkpoint can support multiple point budgets.

At each denoising step, the conditioning input is constructed exactly as during training (Figure~\ref{fig:pipeline}), using the input image and the current noisy offsets.

We use a truncated SDEdit-style diffusion schedule. The initial OT offsets are forward-noised to $t=500$, corresponding to 50\% of the full $T=1000$-step schedule, and standard reverse DDPM sampling is then performed from $t=500$ to $t=0$. The resulting OT offsets are finally converted back to continuous point coordinates to obtain the stipple pattern (Figure~\ref{fig:pipeline}).

\begin{figure*}[t]
    \centering
    \includegraphics[width=0.9\linewidth]{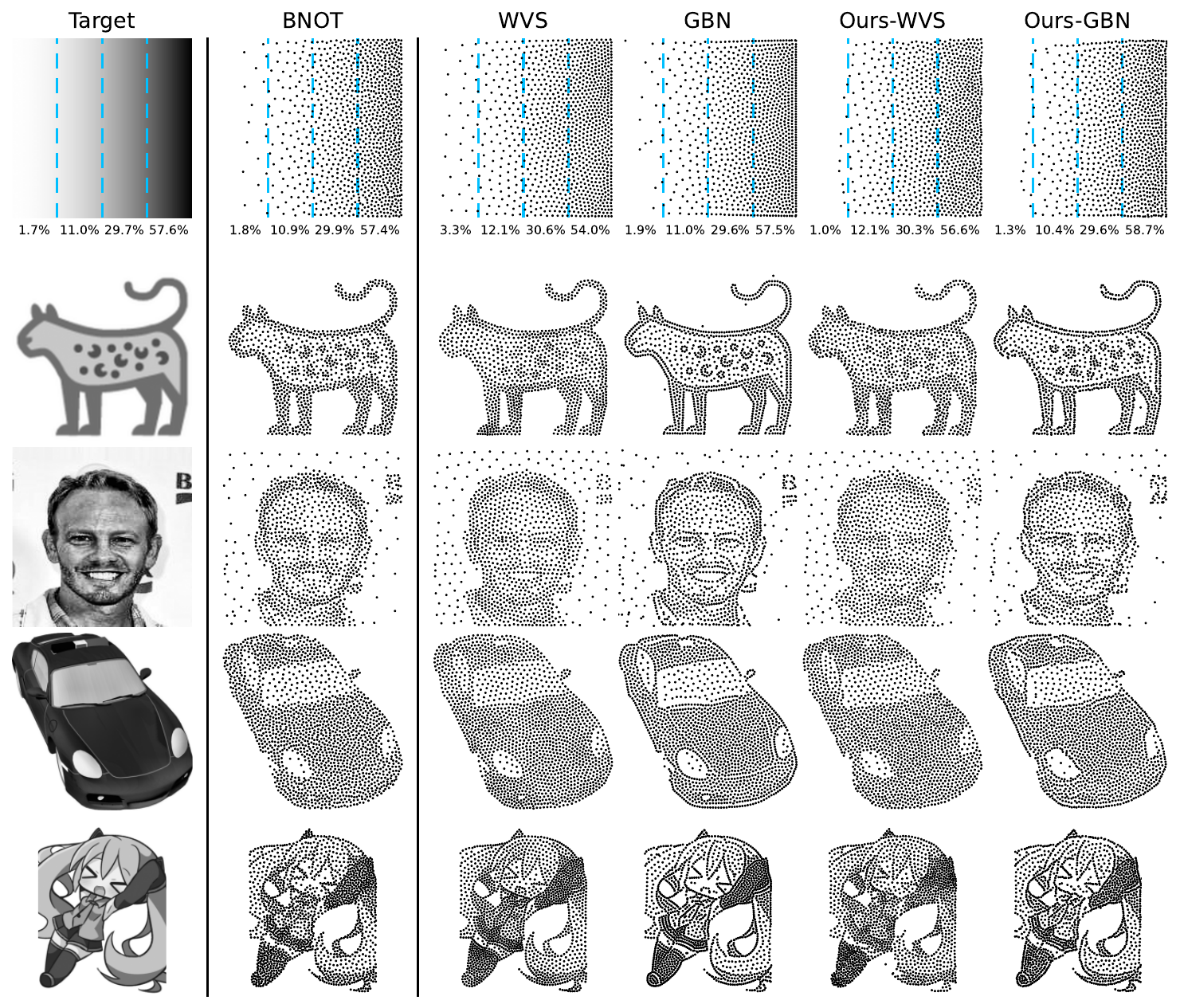}
    \caption{Qualitative comparison on diverse target images with the two exemplar samplers and the BNOT reference. Each row shows the target image followed by stipple patterns generated by BNOT, WVS, GBN, Ours-WVS, and Ours-GBN. Results are shown for a gradient image (top row), with the prescribed point mass per strip below each panel, as well as images from the Icons-50, CelebA, ShapeNet2D, and Anime Chibi datasets respectively.}
    \label{fig:qualitative}
\end{figure*}

\section{Experiments}
\label{sec:experiments}

\paragraph{Datasets}

We evaluate our method on four datasets, converted to grayscale and paired with offline exemplar stipples generated by Weighted Voronoi Stippling (WVS)~\cite{10.1145/508530.508537} and Gaussian Blue Noise (GBN)~\cite{Ahmed2022Gaussian}. Each exemplar uses the point budget of the corresponding OT grid.

\textbf{Icons-50}~\cite{hendrycks2018icons50} is our first benchmark, comprising all $10$K images split into $9$K/$1$K for training/testing, resized to $512\times512$, with $1024$ points on a $32\times32$ grid. \textbf{CelebA}~\cite{liu2015celeba} provides a natural-image benchmark using $5$K images split $4.5$K/$0.5$K, at the same resolution and point budget. The last two benchmarks are non-square, use $1600$ points on a $40\times40$ grid, and hold out $10\%$ of their images for testing. \textbf{ShapeNet2D} contains $4$K renders of ShapeNet~\cite{chang2015shapenetinformationrich3dmodel} objects from seven categories (cabinet, car, sofa, jar, trash bin, bottle, and bed) with the longer side at $1024$ pixels, split $3.6$K/$0.4$K; its renders are color-filtered before the exemplars are generated. \textbf{Anime Chibi}~\cite{AnimeChibiKaggle} contains $2{,}634$ chibi illustrations, resized to a longer side of at most $1024$ pixels with the aspect ratio kept, and split $2{,}371$/$263$.

Each dataset has its own point budget and image resolution. For each dataset we train two checkpoints, one per exemplar sampler (WVS and GBN), at the grid size set by its budget, eight checkpoints in total. Each is trained from scratch for $5000$ epochs with batch size $16$ and learning rate $10^{-4}$. All share the same architecture and differ only in their training exemplars, which isolates the effect of the imitated sampler. Each benchmark is evaluated with its own checkpoints on its held-out split.

\paragraph{Qualitative Results}

Figure~\ref{fig:teaser} illustrates capacity-constrained stippling with our diffusion model on diverse target images. The controlled denoising process progressively refines the initial point distribution while preserving local point repulsion and the prescribed density (left). Our resulting stipple patterns accurately capture the target shapes and spatial density across six target images (right).

Figure~\ref{fig:qualitative} compares each learned variant with the sampler that produced its exemplars, WVS and GBN, on different targets, with BNOT as a reference. Further held-out examples are shown in Appendix~\ref{app:qualitative}.

In the top row, we evaluate the stippling distribution on a continuous quadratic density gradient, $\rho(x)=x^2+\tfrac{1}{500}$ on $[0,1]^2$, partitioned into four equal-width vertical strips with target point masses of 1.7\%, 11.0\%, 29.7\%, and 57.6\%.

We compare our model trained on Icons-50 with the WVS and GBN samplers. Both Ours-WVS and Ours-GBN produce visually comparable results across density levels and closely match the prescribed point-capacity percentages. Our 1024-point outputs closely track the target strip masses, staying within about one percentage point of each target, while using one model per exemplar sampler, without per-target optimization. Both learned variants slightly under-fill the sparsest strip (Ours-WVS 1.0\%, Ours-GBN 1.3\%, target 1.7\%), and Ours-WVS is closer to the target than WVS in the first and the last strip. The density-weighted initialization allocates the mass per region, and the denoising preserves it to within about one percentage point.

The remaining rows show results for target images from the Icons-50, CelebA, ShapeNet2D, and Anime Chibi test sets, respectively. Across these examples, our method preserves blue-noise-like point spacing while closely following the target density and image structure, using one checkpoint per dataset and exemplar sampler.

\begin{figure}[t]
    \centering
    \includegraphics[width=\linewidth]{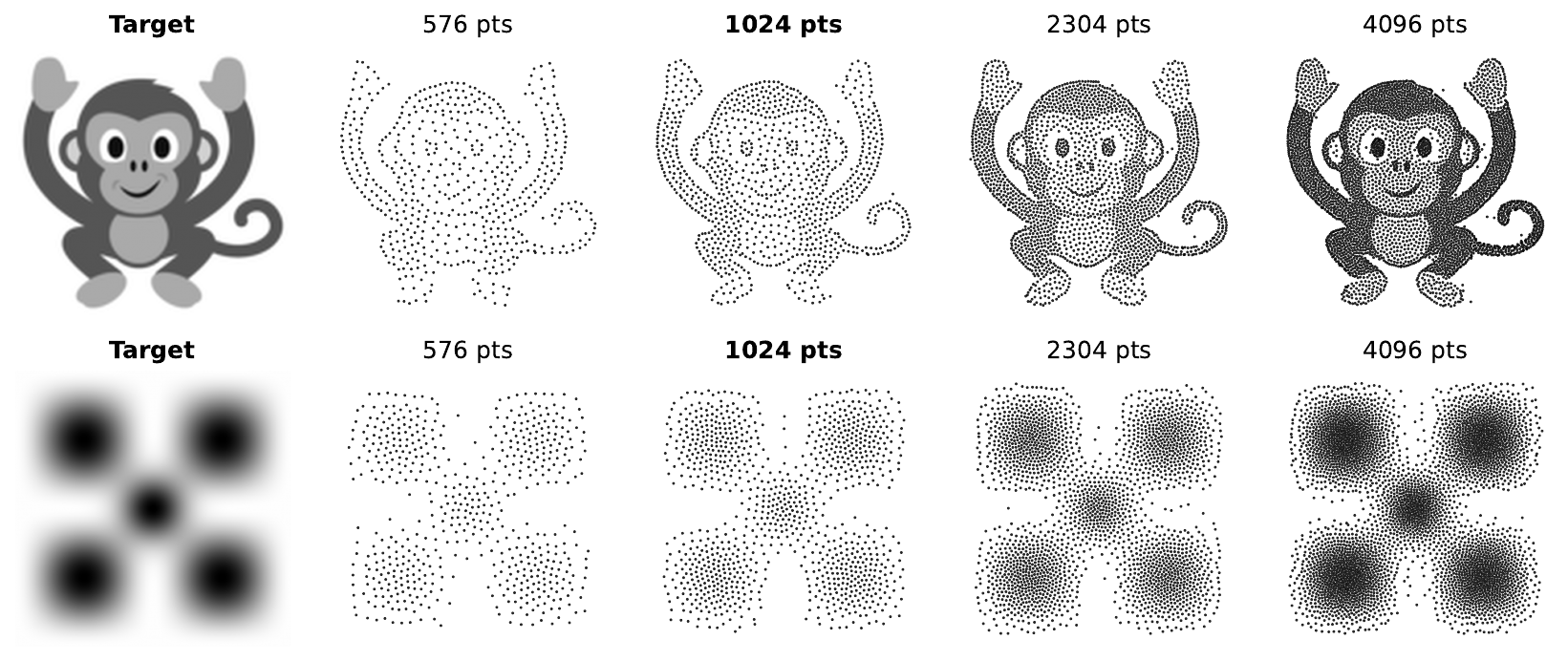}
    \caption{
    Capacity generalization. StippleDiffusion is trained with a 1024-point budget but supports varying point counts at inference. For each target density, we show results with 576, 1024, 2304, and 4096 points, demonstrating that the learned conditional sampler preserves the target density and spatial structure across different output capacities.}
    \label{fig:point_budget_generalization}
\end{figure}

\paragraph{Capacity Generalization}

Although our Icons-50 model is trained exclusively on 1024-point sets ($32\times32$ grid), the OT-grid representation allows inference at different grid resolutions, with output capacities restricted to square numbers of points. Figure~\ref{fig:point_budget_generalization} shows results for two target densities using 576, 1024, 2304, and 4096 points.

Across capacities, the generated patterns preserve the target density and structure, with higher budgets progressively recovering finer details. Thus, a single checkpoint generalizes across substantially different output capacities without retraining, unlike optimization-based methods that require separate optimization for each point budget. At 4,096 points, the icon stays clean, while the smooth density becomes mottled in its falloff zones, indicating the practical limit of capacity extrapolation beyond the training resolution.

\begin{figure}[t]
    \centering
    \includegraphics[width=\linewidth]{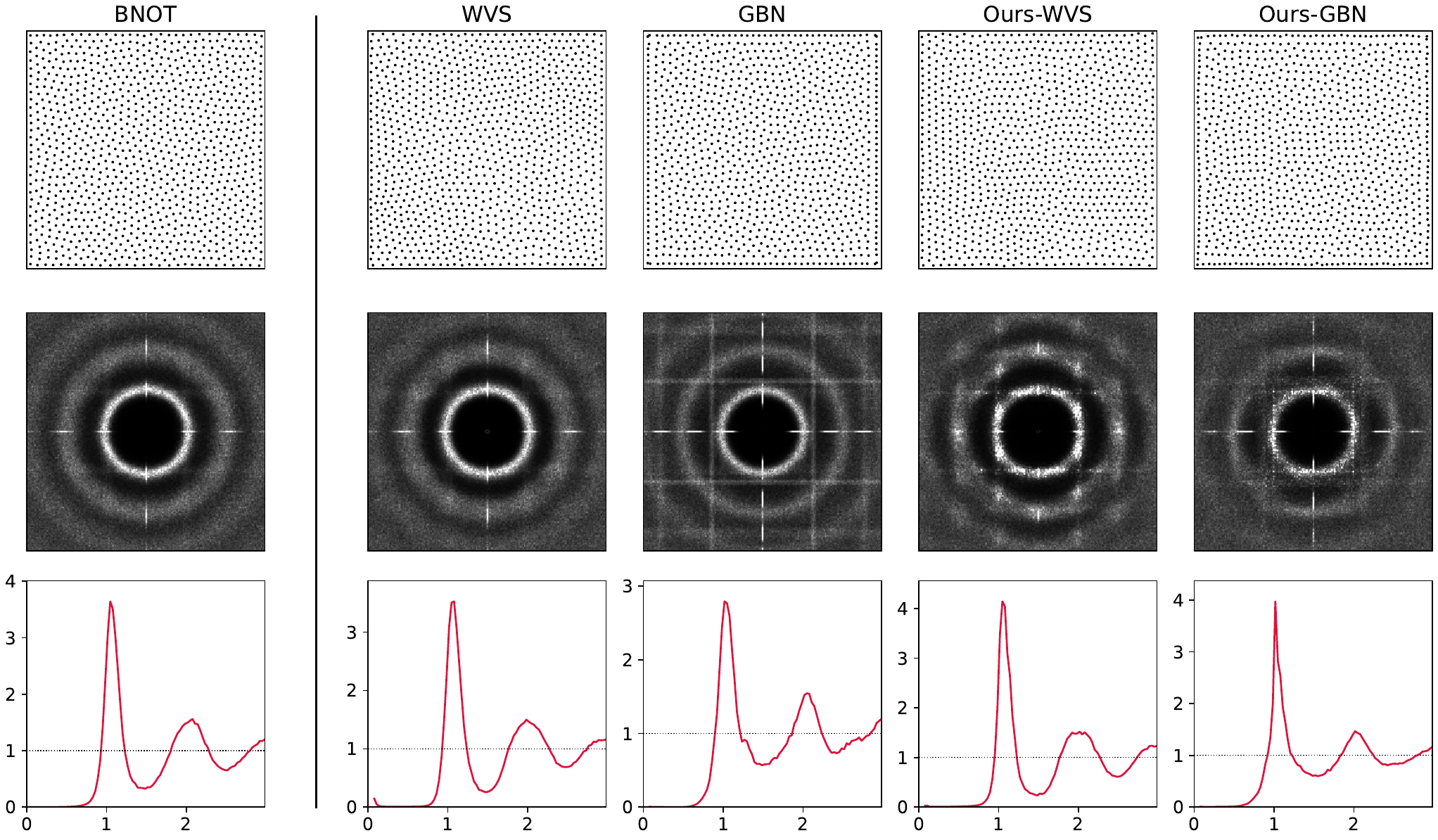}
    \caption{Power spectra for a uniform target ($\rho=1.00$) with 1024 points, averaged over 25 realizations. Top: representative stippling. Middle: mean 2D spectra. Bottom: radial averages. Blue-noise patterns exhibit low-frequency suppression, a peak near the principal frequency, and a flat unit tail; flat spectra indicate white noise, while sharp peaks and spiky tails indicate lattice structure. The radial spectra of our variants closely resemble those of the optimization-based methods.}
    \label{fig:Spectral3}
\end{figure}

\paragraph{Spectral Analysis}

Blue-noise quality is commonly characterized by capacity adherence and spatial regularity, as well as by its frequency-domain signature. Because capacity-constrained stipples are inherently non-stationary, spectra of natural images are dominated by their target structure. We therefore evaluate a constant-density target ($\rho=1.00$) using 1024 points and 25 realizations. Figure~\ref{fig:Spectral3} shows the resulting spectra; pair correlations at several densities are provided in Appendix~\ref{app:spectral}.

All methods exhibit a principal blue-noise peak at approximately $1.05$. Our models show spectral characteristics comparable to those of the optimization-based samplers, with Ours-GBN producing the cleanest high-frequency tail. The optimization-based samplers achieve lower low-frequency floors, indicating stronger suppression of low-frequency energy. The 2D spectra of the learned variants show a mild four-fold modulation aligned with the grid axes, which the radial average hides. We attribute it to the grid representation. Overall, our learned sampler preserves the characteristic blue-noise spectral structure, with modest differences in its frequency response.

\begin{figure}[t]
    \centering
    \includegraphics[width=\linewidth]{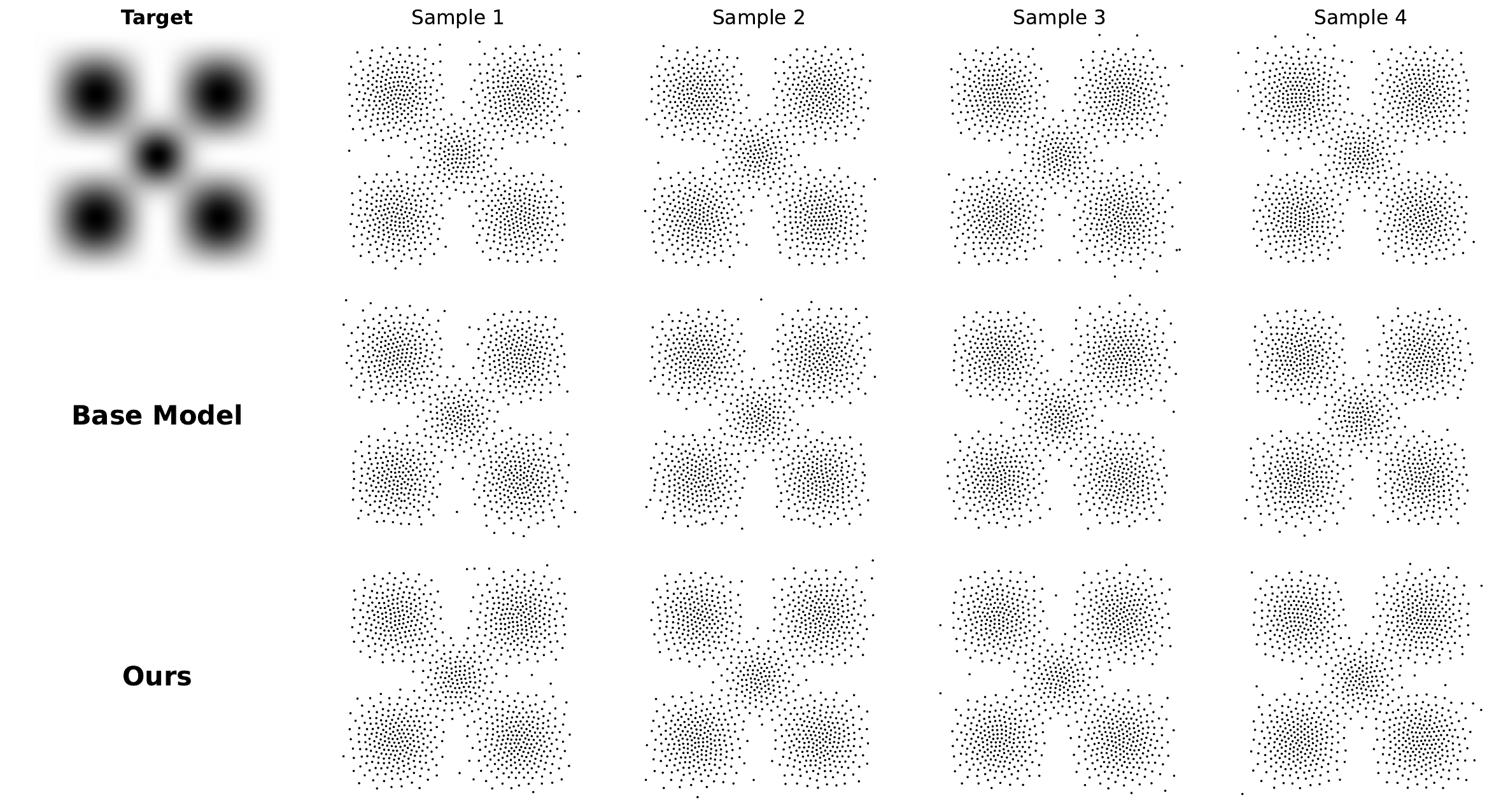}\hfill
    \includegraphics[width=\linewidth]{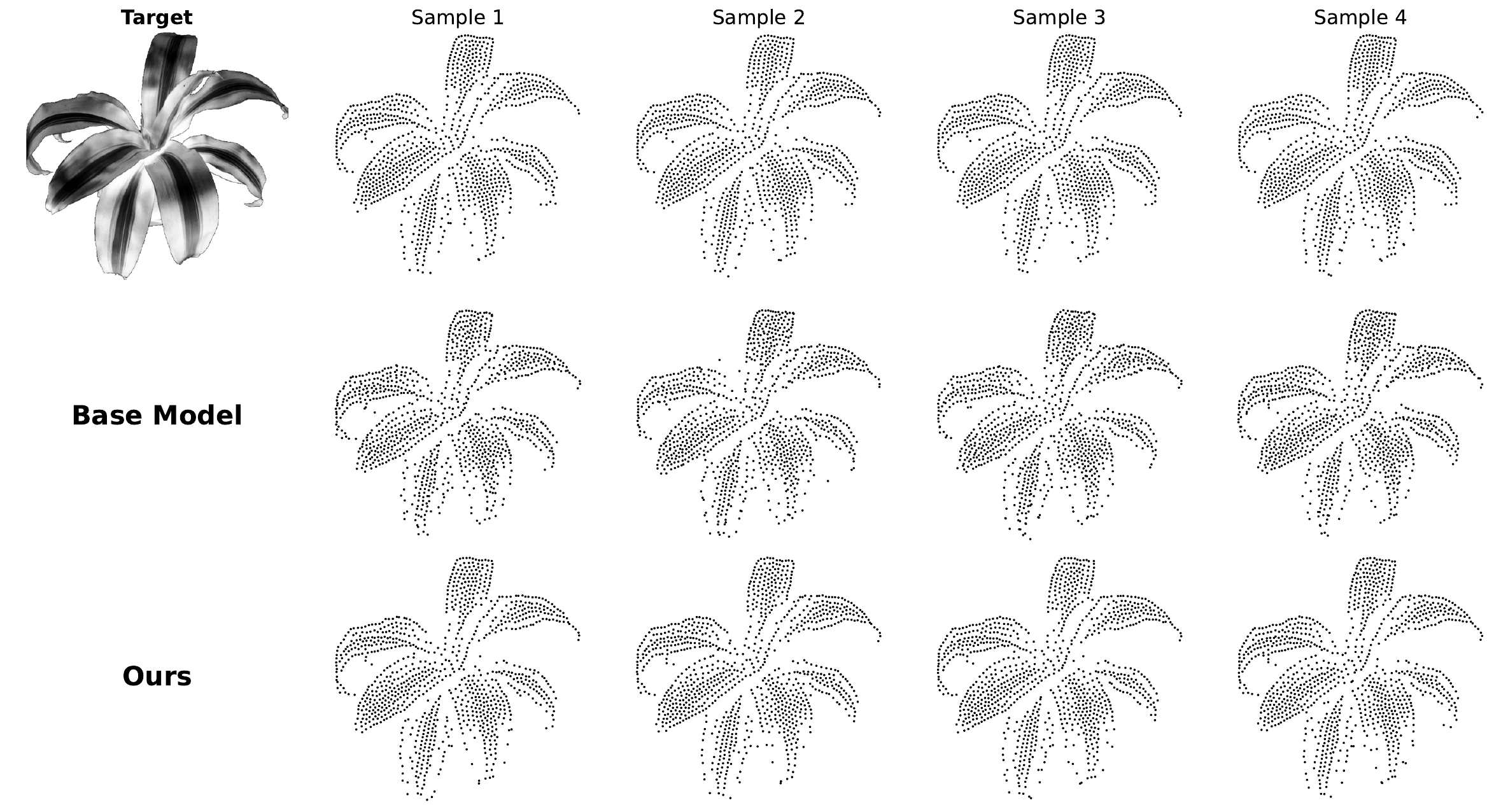}
    \caption{Comparison with fixed-target OT diffusion. Top: A target distribution with $1024$ points and $256$ GBN exemplars (first row: four of them), used to train the fixed-target OT diffusion model (base model). Both models reproduce the target density while preserving the GBN characteristics (second and third rows). Bottom: Image-based stippling with $1024$ points, using $1024$ GBN exemplars generated from the target image to train the base model. Our conditioned sampler follows the image structure while maintaining a well-distributed pattern.
}
    \label{fig:Stress}
\end{figure}

\paragraph{Comparison with Fixed-Target Base Model}
The optimal-transport diffusion model of~\citet{10.1145/3610548.3618243} (i.e., base model) can generate non-uniform point distributions when trained on examples from a fixed target density. In Figure~\ref{fig:Stress}, we qualitatively compare the base model with our conditioned model and verify that the conditioning branch preserves its ability to generate multiple samples from the same target distribution.

In the top experiment, the target density is
$
\rho(x,y)=0.2e^{-20(x^2+y^2)}
+0.2\sin^2(\pi x)\sin^2(\pi y)
$
~\cite{10.1145/1576246.1531392}. This density is challenging due to its sharp localized wells, requiring both accurate global mass allocation and strong local point repulsion. We generate a 1024-point distribution by importance sampling from this density and use the resulting point set to generate GBN exemplars for training.

When trained on the same GBN exemplars as the base model, our model captures the main density structure while maintaining regular point spacing. Multiple generated samples exhibit consistent density structure and point spacing, demonstrating that the conditioning branch preserves the stochastic generation capability of the underlying diffusion model.

The bottom example extends this experiment to a natural image, again using GBN exemplars generated from the image to train the base model.
 Our generated point sets follow the image structure while maintaining a well-distributed pattern. Together, these results show that the conditioning branch preserves the base model's ability to reproduce the prescribed exemplar distribution.

\paragraph{Metrics}
\label{sec:metrics}

We evaluate each point set using seven metrics grouped by the property they assess: relaxation (CVT, HOT), capacity (capacity deviation, centroid displacement), global agreement with the target density (EMD, Sinkhorn), and local spacing (spatial measure).

\begin{itemize}

\item \textbf{CVT energy}~\cite{Du1999CVT}. Measures point relaxation by integrating, over each point's Voronoi cell, the density-weighted squared distance from every location in the cell to that point. It is minimized when each point coincides with the density-weighted centroid of its cell, as in Lloyd-based stippling~\cite{10.1145/508530.508537}:
$$
E_{CVT}=\sum_{i=1}^{n}\int_{V_i}\rho(x)\,\|x-s_i\|^2\,dx ,
$$
where $s_i$ is the $i$-th point, $V_i$ its Voronoi cell, and $\rho$ the target density.

\item \textbf{HOT energy}~\cite{Mullen2011HOT}. The analogous energy over power cells, whose weights are chosen so that all cells have equal mass~\cite{Aurenhammer1987Power}. This is the $\star_0$-HOT$_{2,2}$ energy minimized by BNOT~\cite{deGoes:2012:BNOT}:
$$
E_{HOT}=\sum_{i=1}^{n}\int_{V^w_i}\rho(x)\,\|x-s_i\|^2\,dx ,
$$
where $V^w_i$ is the power cell of $s_i$.

\item \textbf{Capacity deviation $\delta_c$}~\cite{10.1145/1576246.1531392}. Measures how uniformly the target mass is distributed across Voronoi cells. It is zero when every cell contains the same mass:
$$
\delta_c=\frac{1}{n}\sum_{i=1}^{n}\left(\frac{m_i}{\bar{m}}-1\right)^2 ,
$$
where $m_i=\int_{V_i}\rho(x)\,dx$ and $\bar{m}$ is the mean cell mass.

\item \textbf{Centroid displacement $D$}. Measures the distance between each point and the density-weighted centroid of its power cell. It is zero when the BNOT centroidal optimality condition is satisfied~\cite{deGoes:2012:BNOT}:
$$
D=\frac{1}{n}\sum_{i=1}^{n}\|s_i-b_i\|,\qquad
b_i=\frac{\int_{V^w_i}x\,\rho(x)\,dx}{\int_{V^w_i}\rho(x)\,dx}.
$$
\item \textbf{EMD and Sinkhorn}. Measure global agreement with the target density by transporting the point distribution to the target distribution. EMD is the 2-Wasserstein distance~\cite{Villani2009OT}; Sinkhorn is its entropy-regularized approximation~\cite{cuturi2013sinkhorndistanceslightspeedcomputation}. Both measure global density agreement without directly assessing local point arrangement.

\item \textbf{Spatial measure $\rho_s$}~\cite{lagae2008}. Measures local point spacing after rescaling distances by the local density~\cite{10.1145/2010324.1964945}. Higher values indicate more uniform nearest-neighbor spacing relative to the local density:
$$
\rho_s=\frac{1}{n}\sum_{i=1}^{n}
\frac{r_i\sqrt{n\,\rho(s_i)/\bar{\rho}}}{r_{\max}},
$$
where $r_i$ is the nearest-neighbor distance of $s_i$, $\bar{\rho}$ is the mean target density at the points, and $r_{\max}=(2/\sqrt{3})^{1/2}$ is the reference spacing for hexagonal packing at unit density.

\end{itemize}

\paragraph{Quantitative Comparison}

\begin{table}[t]
\centering
\small
\setlength{\tabcolsep}{4pt}
\renewcommand{\arraystretch}{1.200}
\begin{tabular}{|l|c|c|c|c|}
\hline
\textbf{Metric} & WVS & GBN & Ours-WVS & Ours-GBN \\
\hline
\multicolumn{5}{|l|}{\textbf{Icons-50 (1000 images, 1024 points)}} \\
\hline
CVT ($\times 10^{-5}$)\,$\downarrow$ & \textbf{2.194} & 2.370 & \underline{2.265} & 2.450 \\
HOT ($\times 10^{-5}$)\,$\downarrow$ & 10.536 & \underline{5.204} & 6.459 & \textbf{4.059} \\
Capacity $\delta_c$\,$\downarrow$ & 0.041 & \textbf{0.010} & 0.041 & \underline{0.018} \\
Centroid $D$\,$\downarrow$ & 0.014 & \underline{0.008} & 0.010 & \textbf{0.006} \\
EMD\,$\downarrow$ & 0.020 & \underline{0.015} & 0.016 & \textbf{0.013} \\
Sinkhorn\,$\downarrow$ & 0.094 & \underline{0.093} & \underline{0.093} & \textbf{0.092} \\
Spatial $\rho_s$\,$\uparrow$ & \textbf{0.589} & 0.516 & \underline{0.562} & 0.494 \\
\hline
\multicolumn{5}{|l|}{\textbf{CelebA (500 images, 1024 points)}} \\
\hline
CVT ($\times 10^{-5}$)\,$\downarrow$ & \textbf{5.249} & 6.357 & \underline{5.687} & 6.766 \\
HOT ($\times 10^{-5}$)\,$\downarrow$ & 21.807 & 12.864 & \underline{11.030} & \textbf{8.265} \\
Capacity $\delta_c$\,$\downarrow$ & 0.054 & \textbf{0.013} & 0.051 & \underline{0.022} \\
Centroid $D$\,$\downarrow$ & 0.016 & 0.010 & \underline{0.009} & \textbf{0.005} \\
EMD\,$\downarrow$ & 0.024 & 0.018 & \underline{0.017} & \textbf{0.015} \\
Sinkhorn\,$\downarrow$ & 0.097 & \underline{0.096} & \underline{0.096} & \textbf{0.095} \\
Spatial $\rho_s$\,$\uparrow$ & \textbf{0.741} & 0.604 & \underline{0.718} & 0.599 \\
\hline
\multicolumn{5}{|l|}{\textbf{ShapeNet2D (400 images, 1600 points)}} \\
\hline
CVT ($\times 10^{-5}$)\,$\downarrow$ & \textbf{2.629} & \underline{3.391} & 3.530 & 3.688 \\
HOT ($\times 10^{-5}$)\,$\downarrow$ & 11.432 & \underline{7.055} & 7.253 & \textbf{5.367} \\
Capacity $\delta_c$\,$\downarrow$ & 0.043 & \textbf{0.013} & 0.059 & \underline{0.038} \\
Centroid $D$\,$\downarrow$ & 0.012 & \underline{0.007} & 0.008 & \textbf{0.006} \\
EMD\,$\downarrow$ & 0.019 & \underline{0.015} & \underline{0.015} & \textbf{0.013} \\
Sinkhorn\,$\downarrow$ & \underline{0.095} & \textbf{0.094} & \underline{0.095} & \textbf{0.094} \\
Spatial $\rho_s$\,$\uparrow$ & \textbf{0.600} & 0.534 & \underline{0.574} & 0.521 \\
\hline
\multicolumn{5}{|l|}{\textbf{Anime Chibi (263 images, 1600 points)}} \\
\hline
CVT ($\times 10^{-5}$)\,$\downarrow$ & \textbf{0.738} & \underline{0.913} & 1.146 & 1.307 \\
HOT ($\times 10^{-5}$)\,$\downarrow$ & 4.655 & 3.262 & \underline{3.089} & \textbf{2.608} \\
Capacity $\delta_c$\,$\downarrow$ & \underline{0.061} & \textbf{0.030} & 0.129 & 0.107 \\
Centroid $D$\,$\downarrow$ & 0.011 & 0.010 & \underline{0.009} & \textbf{0.007} \\
EMD\,$\downarrow$ & 0.017 & 0.016 & \underline{0.014} & \textbf{0.013} \\
Sinkhorn\,$\downarrow$ & \underline{0.094} & \underline{0.094} & \textbf{0.093} & \textbf{0.093} \\
Spatial $\rho_s$\,$\uparrow$ & \textbf{0.469} & 0.391 & \underline{0.433} & 0.368 \\
\hline
\end{tabular}
\caption{Quantitative comparison of WVS, GBN, and our two variants, Ours-WVS and Ours-GBN, trained on WVS and GBN exemplars respectively, across four benchmarks. All values are means over the respective test sets at each benchmark's point budget; best and second-best results per row are shown in bold and underlined, respectively. Each benchmark uses its corresponding trained checkpoints.}
\label{tab:quantitative_comparison}
\end{table}

In Table~\ref{tab:quantitative_comparison}, we compare our model with the optimization-based samplers WVS and GBN on all four datasets: the 1K-image Icons-50 and 500-image CelebA test splits at 1024 points, and the 400-image ShapeNet2D and 263-image Anime Chibi test splits at 1600 points (subtables). Our model is trained separately on WVS and GBN exemplars, and both variants are reported. Bold and underlined values denote the best and second-best results, respectively.

Compared with its own exemplar sampler, each learned variant improves HOT energy, centroid displacement and EMD on all four datasets (on Icons-50, the HOT energy is $4.059$ for Ours-GBN against $5.204$ for GBN), and falls short in CVT energy and spatial measure. The Sinkhorn values of all methods lie within $0.002$ of each other on every dataset. Capacity deviation stays close to the exemplar sampler on the two square 1024-point datasets, but the gap grows on the two non-square 1600-point datasets (Ours-GBN $0.038$ against $0.013$ for GBN on ShapeNet2D, and $0.107$ against $0.030$ on Anime Chibi). These two datasets differ from the first two in point budget, aspect ratio and training-set size at once, and our experiments do not separate these factors.

The learned variants inherit the ranking of their exemplar samplers. In 27 of the 28 dataset--metric pairs, the variant trained on the better of WVS and GBN is also the better of our two variants, and the remaining pair (Sinkhorn on Anime Chibi) is a tie for both. Within the table, our variants rank best or second-best on all but three pairs (CVT on ShapeNet2D and Anime Chibi, and capacity deviation on Anime Chibi), while requiring no optimization at inference. WVS minimizes the CVT energy directly and therefore has the best CVT energy, and with it the best spatial measure. GBN's Gaussian-kernel objective promotes capacity preservation and gives the best capacity deviation.

\begin{figure}[t]
    \centering
    \includegraphics[width=1.0\linewidth]{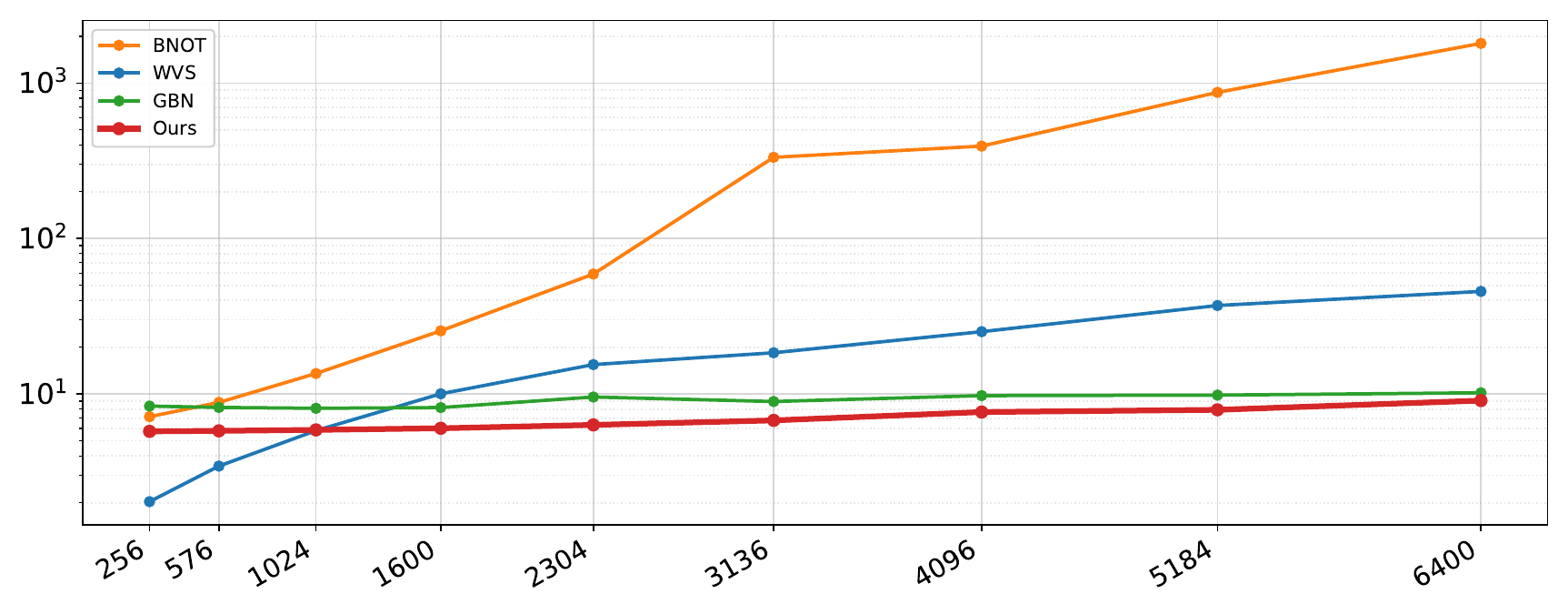}
    \caption{Stippling runtime comparison across point budgets. Mean generation time (seconds) over 10 runs on Icons-50 targets. Ours-GBN and Ours-WVS have identical inference costs and are on par with or faster than all the optimization-based samplers from 1024 points onward.}
    \label{fig:timing}
\end{figure}

\paragraph{Relation to BNOT}
BNOT optimizes the HOT energy directly and sets the reference on the transport metrics. On HOT energy, centroid displacement and EMD, each learned variant lies between its exemplar sampler and BNOT on all four datasets. Ours-WVS has a better spatial measure than BNOT on all four datasets, and a lower CVT energy on the two 1024-point datasets. BNOT reaches its optimum by per-target Newton iterations whose cost grows by more than two orders of magnitude across our budget range (Figure~\ref{fig:timing}). Our variants instead run at a cost that grows little with the budget, generalize to unseen images and budgets, and are differentiable. BNOT appears as an unranked reference in the figures and in Table~\ref{tab:quantitative_comparison_full} (Appendix~\ref{app:full_quantitative}).

\paragraph{Running-time Evaluation}

We report runtime measurements in Figure~\ref{fig:timing} across multiple point budgets. Ours-WVS and Ours-GBN share the same architecture and therefore have identical inference costs. For each method and point count, we report the mean time to generate one stipple pattern.

The optimization-based samplers use their published default stopping criteria: 50 Lloyd iterations for WVS, up to 25 Newton iterations for BNOT, and 1000 optimization steps for GBN. Our runtime is determined by a fixed number of denoising steps, independent of the target.

Our cost grows little with point count. It is slower than WVS at smaller budgets but becomes the fastest method from about 1024 points onward. It is faster than BNOT and GBN at every evaluated budget, with the gap to BNOT widening as BNOT becomes increasingly expensive. Hardware details and CPU/GPU profiling are provided in Appendix~\ref{app:runtime}.

\begin{table}[t]
\centering
\small
\setlength{\tabcolsep}{3pt}
\renewcommand{\arraystretch}{1.05}
\resizebox{\columnwidth}{!}{%
\begin{tabular}{|l||c|c|c|c|c|c|}
\hline
 & Vanilla & Unfrozen & GECCO & Gated & \textbf{Full} & \textbf{SDEdit} \\
\hline
Infer Time (s) $\downarrow$ & 10.38 & 10.40 & 10.96 & 10.77 & 11.08 & \textbf{5.65} \\
\hline
CVT ($\times 10^{-5}$) $\downarrow$ & 2.532 & 2.423 & 2.463 & 2.558 & \textbf{2.399} & 2.412 \\
HOT ($\times 10^{-5}$) $\downarrow$ & 5.170 & 6.444 & 4.819 & 5.001 & \textbf{4.518} & 4.978 \\
Capacity $\delta_c$ $\downarrow$ & 0.0626 & 0.0433 & 0.0412 & 0.0609 & 0.0259 & \textbf{0.0255} \\
Centroid $D$ $\downarrow$ & 0.0089 & 0.0115 & 0.0091 & 0.0103 & 0.0078 & \textbf{0.0071} \\
EMD $\downarrow$ & 0.0170 & 0.0187 & 0.0190 & 0.0163 & 0.0151 & \textbf{0.0146} \\
Sinkhorn $\downarrow$ & 0.0933 & 0.0936 & 0.0949 & 0.0934 & 0.0930 & \textbf{0.0926} \\
Spatial $\rho_s$ $\uparrow$ & 0.465 & \textbf{0.490} & 0.463 & 0.463 & 0.485 & 0.484 \\
\hline
\end{tabular}
}
\caption{Quantitative ablation on Icons-50 using the GBN-trained variant at 1024 points. The five architectural variants are separately trained checkpoints sampled from pure noise over the full diffusion schedule; \textbf{SDEdit} samples the \textbf{Full} checkpoint from a density-weighted rejection-sampled prior over the final 50\% of the schedule. All variants use the same base U-Net, Min-SNR-$\gamma=5.0$ weighting, and KDE loss, and are trained for 5,000 epochs on a $32\times32$ grid. Best values per row are shown in bold.
}
\label{table:ablations}
\end{table}

\paragraph{Ablation Study}

For the architectural ablations, we disable all inference-time modifications and use standard diffusion sampling from pure noise. We compare our full model with four simplified variants: \emph{Vanilla}, a standard ControlNet applied to the frozen base model of \citet{10.1145/3610548.3618243}; \emph{Unfrozen}, which jointly trains the randomly initialized base U-Net and control branch; \emph{GECCO}, which augments the frozen Vanilla model with continuous GECCO feature extraction; and \emph{Gated}, which replaces zero-convolutions with transform-gated $1\times1$ projections. The \emph{Full} model combines joint training with both GECCO and Gated augmentations.

\begin{figure}[ht]
    \centering
    \includegraphics[width=1.0\linewidth]{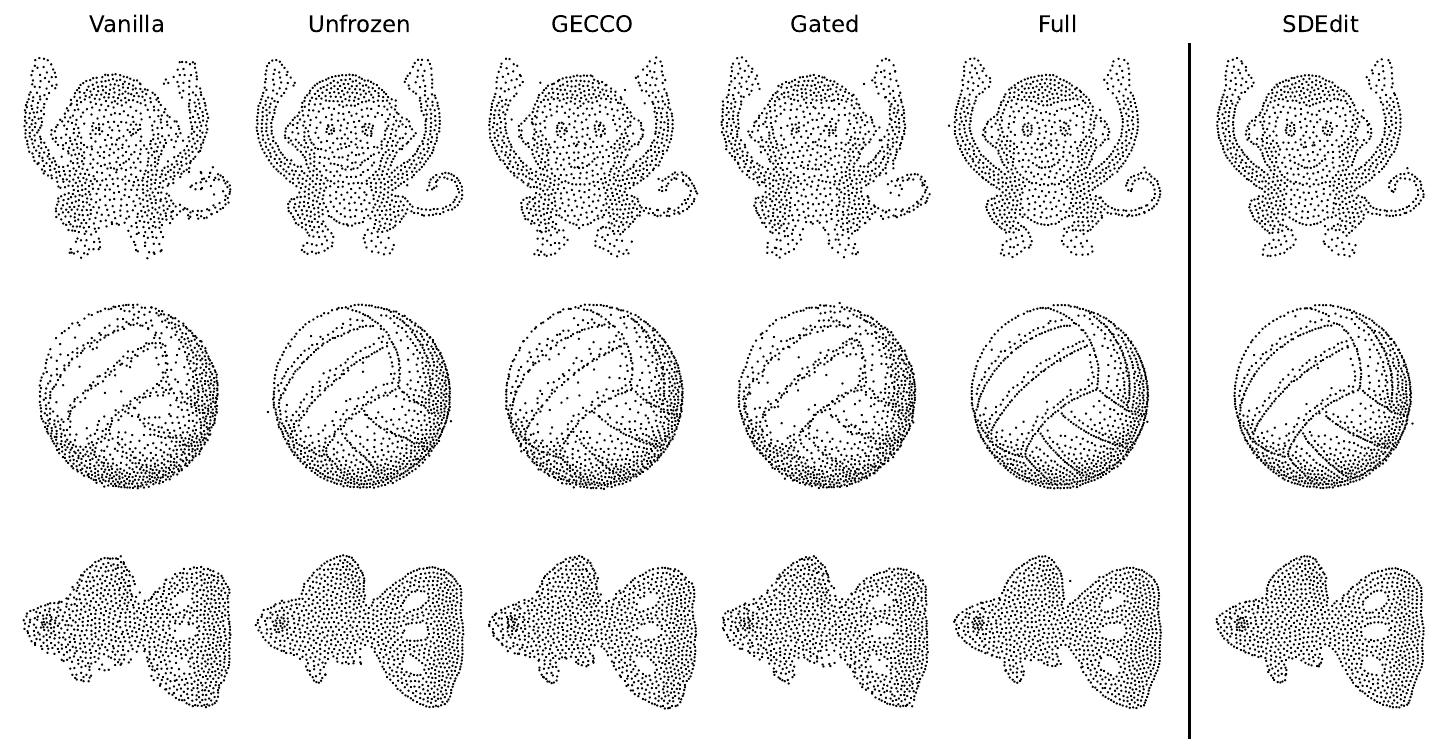}
    \caption{Ablation study on three target images (rows), comparing six model variants (columns):
    \textit{Vanilla}, \textit{Unfrozen}, \textit{GECCO}, \textit{Gated}, \textit{Full}, and \textit{SDEdit}.
    }
    \label{fig:ablation_visual}
\end{figure}

The architectural ablation results are summarized in Table~\ref{table:ablations}, with qualitative comparisons shown in Figure~\ref{fig:ablation_visual}. \emph{Full} lowers capacity deviation from $0.0626$ (\emph{Vanilla}) to $0.0259$ and also improves HOT energy, EMD and centroid displacement. Gating alone brings little over \emph{Vanilla}; the gains come from joint training and the per-point features, and are largest when the two are combined. Figure~\ref{fig:ablation_visual} shows the same ordering: \emph{Vanilla} and \emph{Gated} leave uneven point density inside the contours, while \emph{Full} is close to the final \emph{SDEdit} sampler.

We next evaluate the inference-time sampling strategy using the \emph{Full} model. Following SDEdit~\cite{meng2022sdeditguidedimagesynthesis}, we replace pure-noise initialization with a density-weighted rejection-sampled prior and denoise only the final 50\% of the diffusion schedule.

As shown in Table~\ref{table:ablations}, inference time decreases from $11.08$\,s to $5.65$\,s at comparable quality. Capacity deviation is essentially unchanged ($0.0255$ against $0.0259$), centroid displacement and EMD improve slightly, and HOT energy is slightly worse. Qualitatively, \emph{SDEdit} and \emph{Full} produce similar results (Figure~\ref{fig:ablation_visual}), and both look clearly better than \emph{Vanilla}.

\begin{figure*}[t]
    \centering
    \includegraphics[width=1.0\linewidth]{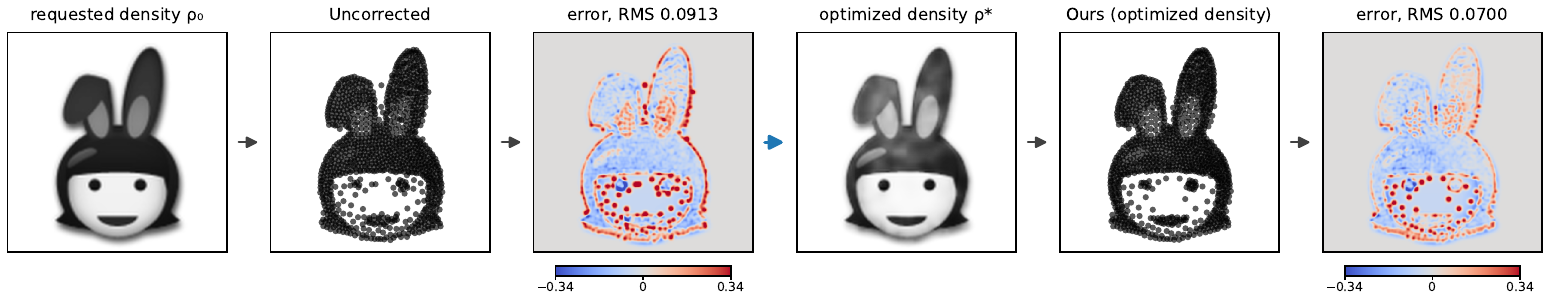}
    \caption{Differentiable tone correction. Left: requesting the target density $\rho_0$ produces a stipple whose rendered tone deviates from the target (error map: red = too dark, blue = too light). The blue arrow denotes our optimization of the requested density through the differentiable sampler. Right: the optimized density $\rho^\star$, its rendered stipple, and the resulting reduced error. }
    \label{fig:ToneMechanism}
\end{figure*}

\begin{figure}[t]
    \centering
    \includegraphics[width=1.0\linewidth]{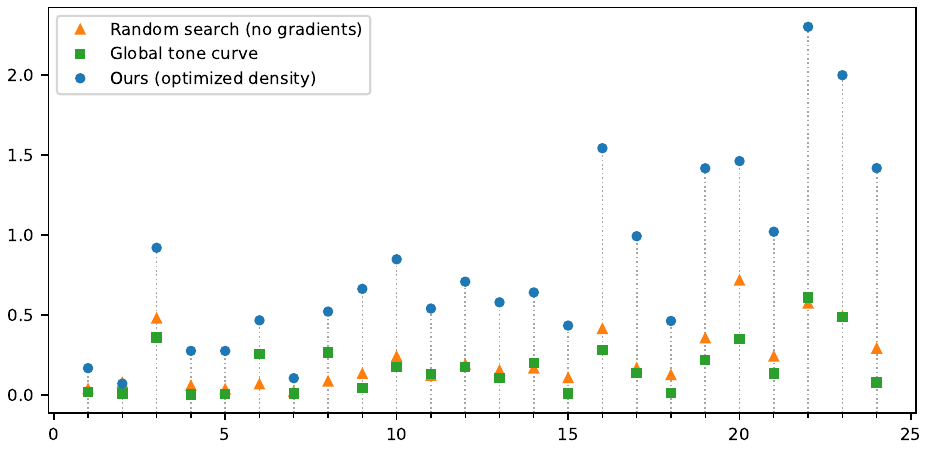}
    \caption{
    Tone correction across 24 target images (x-axis). Per-image PSNR gain over the uncorrected stipple (y-axis, dB) of \textsc{Ours}, \textsc{Random Search}, and \textsc{Tone Curve}. Our gradient-based optimization improves every image, achieving a higher gain than the tone curve on all 24 images.
    }
    \label{fig:ToneGraph}
\end{figure}

\paragraph{Differentiable Tone Correction}

Because our sampler is a neural network, it defines a differentiable mapping $S$ from a density map $\rho$ to a point set. Thus, any differentiable objective on the generated stipple can be optimized with respect to the requested density by back-propagating through $S$. In contrast, the iterative optimization samplers provide no direct gradient; differentiating through WVS, BNOT, or GBN would require back-propagation through their respective Lloyd, Newton, or gradient-based iterations, or implicit differentiation at their fixed point, which we do not attempt here.

We evaluate this capability on dot gain, where finite-sized stipple dots cause the rendered tone to differ from the requested density: isolated dots can appear too dark, while overlapping dots in dark regions saturate (Figure~\ref{fig:ToneMechanism}, left). We therefore optimize the requested density $\rho$ so that the rendered stipple better matches the target darkness.

The renderer $R$ places each point as a Gaussian dot with standard deviation $2$ pixels on a $256\times256$ canvas and converts accumulated ink $a$ to darkness using Beer--Lambert saturation, $1-e^{-\kappa a}$. We calibrate the ink gain $\kappa$ once on six images so that uncorrected stipples match their mean target darkness, and then keep it fixed. Given target image $I$, the target darkness, which is also its target density, is $\rho_0=1-I$. We parameterize the corrected density as
$
\rho=\mathrm{sigmoid}\!\left(\mathrm{logit}(\rho_0)+\phi\right),
$
where $\phi$ is a $64\times64$ correction field bilinearly upsampled to the image resolution. We optimize
$$
\phi^\star=\arg\min_{\phi}\;
\left\lVert R\bigl(S(\rho)\bigr)-\rho_0\right\rVert_2^2
+\lambda_2\lVert\phi\rVert_2^2
+\lambda_{\mathrm{tv}}\mathrm{TV}(\phi),
$$
with $\lambda_2=10^{-3}$ and $\lambda_{\mathrm{tv}}=10^{-2}$. Adam is used with a learning rate of $0.05$ for 300 steps. Only $\rho$ is optimized; the network weights, remaining conditioning, initial points, and noise are fixed, making the objective deterministic.

We validate the gradients using central finite differences with step $0.02$ at 24 randomly selected entries of $\phi$. The analytic and numerical gradients achieve a cosine similarity of $0.995$, agree in sign at all 24 locations, and have a magnitude slope of $0.98$.

We compare four strategies on 24 held-out Icons-50 targets at 1024 points. \textsc{uncorrected} directly uses $\rho_0$. \textsc{tone curve} replaces $\phi$ with a global 17-knot transfer curve and is optimized using the same objective. \textsc{random search} optimizes the full $64\times64$ field using gradient-free $(1{+}1)$ random search with the $1/5$ success rule and 600 sampler evaluations, matching the computational cost of 300 forward-and-backward steps. \textsc{ours} performs the gradient-based optimization described above. We evaluate the rendered stipples using PSNR, SSIM, and mean absolute error (MAE).

For efficiency, we compute gradients through the closed-form $\hat{x}_0$ estimate after a single denoising step at $t=120$, which is already differentiated through during training. Full back-propagation through the reverse process is also possible, with memory proportional to the number of denoising steps. Although the dot-gain model is a simplified printer proxy, the results consistently show improvement across all targets and over gradient-free optimization at equal cost.

Our correction improves all 24 targets, with an average PSNR gain of $+0.83$\,dB, compared with $+0.22$\,dB for \textsc{random search} at equal cost and $+0.17$\,dB for the tone curve (Table~\ref{tab:tone}, Figure~\ref{fig:ToneGraph}). Gradients outperform random search on 23 of 24 images. The optimized density reduces point crowding (Figure~\ref{fig:ToneMechanism}, right), lowering RMS error from $0.0913$ to $0.0700$. Across a $16\times$ range of ink gains, the PSNR improvement remains between $+0.31$ and $+1.44$\,dB and consistently exceeds the tone curve (Appendix~\ref{app:gain}).

\begin{table}[t]
\centering
\small
\begin{tabular}{lccc}
\toprule
strategy & PSNR\,(dB) & SSIM & MAE \\
\midrule
Uncorrected                & 15.76 & 0.5474 & 0.1158 \\
Global tone curve          & 15.93 & 0.5488 & 0.1140 \\
Random search (no grad.)   & 15.98 & 0.5521 & 0.1132 \\
\textbf{Ours (optimized $\rho$)} & \textbf{16.59} & \textbf{0.5725} & \textbf{0.1079} \\
\bottomrule
\end{tabular}
\caption{Tone correction comparison under various metrics.  PSNR, SSIM, and MAE between the rendered stipples and their target images, averaged over 24 targets. \textsc{Ours} achieves the best performance across all three metrics.
}
\label{tab:tone}
\end{table}

\section{Conclusions and Limitations}

We present a learned stippling algorithm that learns the stippling process itself rather than a single point distribution. A single checkpoint per dataset and exemplar sampler generates high-quality stipple patterns for unseen images and a range of point budgets, with an inference cost that grows little with output size and a fully differentiable mapping from density to stipples. This shifts stippling from a per-target optimization problem to a reusable, image-conditioned generative process. It enables fast generation, capacity generalization, and downstream optimization through the sampler.

Visual quality remains the main limitation. Gaussian Blue Noise produces smoother density transitions and preserves fine details and sharp edges more faithfully, while our spectra retain the blue-noise structure but suppress low frequencies less effectively than optimization-based methods.
Our method is also slower than deterministic low-discrepancy samplers, and its OT-grid representation relies on regular grids and convolutions that do not readily extend beyond 2D, motivating non-grid alternatives.
Nevertheless, differentiability enables capabilities unavailable to these optimizers, as demonstrated by our tone-correction experiment. Future work can optimize through the full reverse process, support richer objectives such as stylization, and use generated stipples in downstream learned models.

\bibliographystyle{abbrvnat}
\bibliography{base}

\clearpage
\appendix

\section{Differentiability: Sensitivity to the Ink Gain}
\label{app:gain}

\begin{figure}[t]
    \centering
    \includegraphics[width=\linewidth]{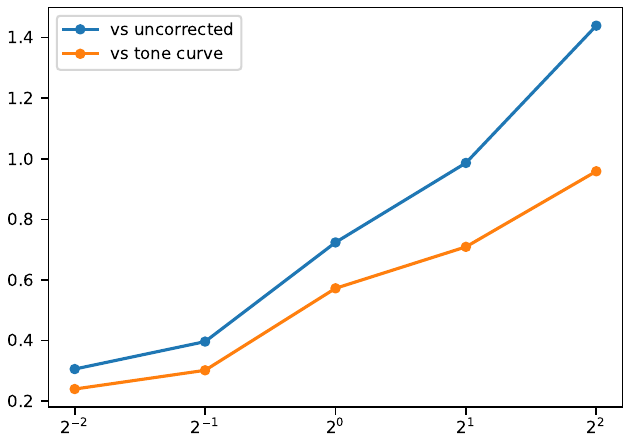}
    \caption{Sensitivity to the rendering model. Ink gain relative to its calibrated value, swept from a quarter to four times (x-axis, log scale), against the PSNR advantage of our optimized density over the uncorrected stipple and over the tone curve (y-axis, dB).}
     \label{fig:sup_gain_sweep}
\end{figure}

Figure~\ref{fig:sup_gain_sweep} sweeps the ink gain of the renderer in the dot-gain experiment of Figure~\ref{fig:ToneMechanism}, to show that the advantage of optimizing through the sampler is not an artifact of the calibrated value.

\section{Runtime Profiling}
\label{app:runtime}

\begin{figure}[t]
    \centering
    \includegraphics[width=\linewidth]{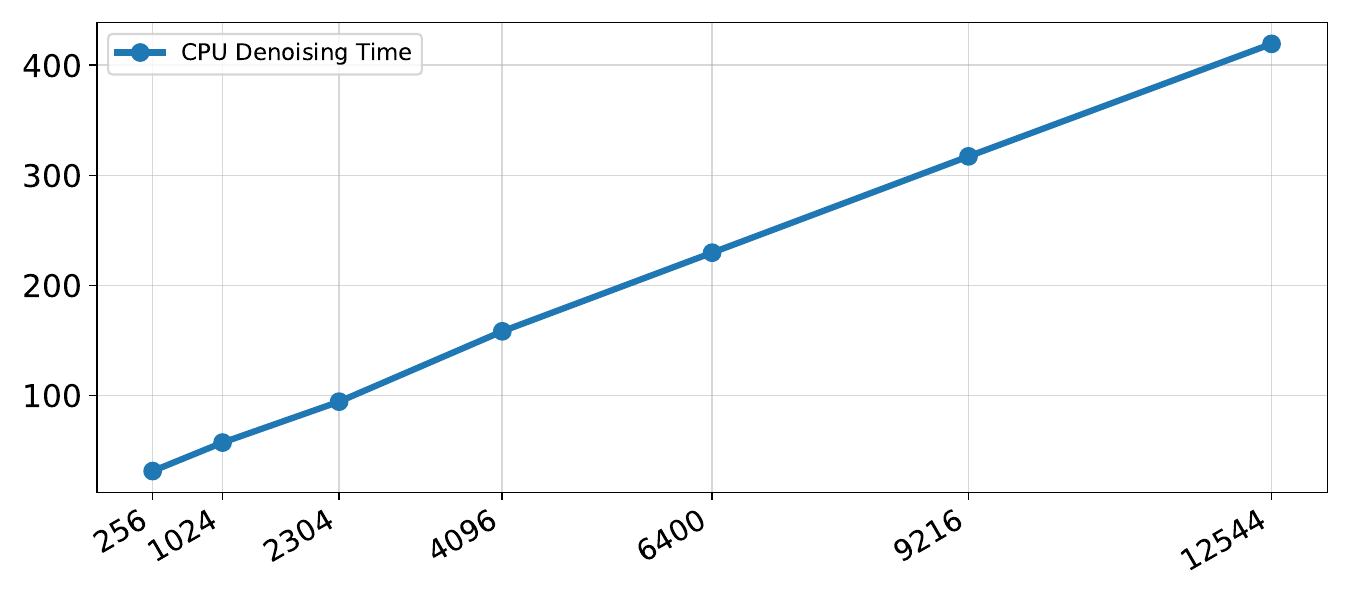}
    \caption{Denoising time on the CPU versus output point count.}
    \label{fig:profiler_times_points1}
\end{figure}

\begin{figure*}[t]
    \centering
    \begin{minipage}{0.49\textwidth}
        \centering
        \includegraphics[width=\linewidth]{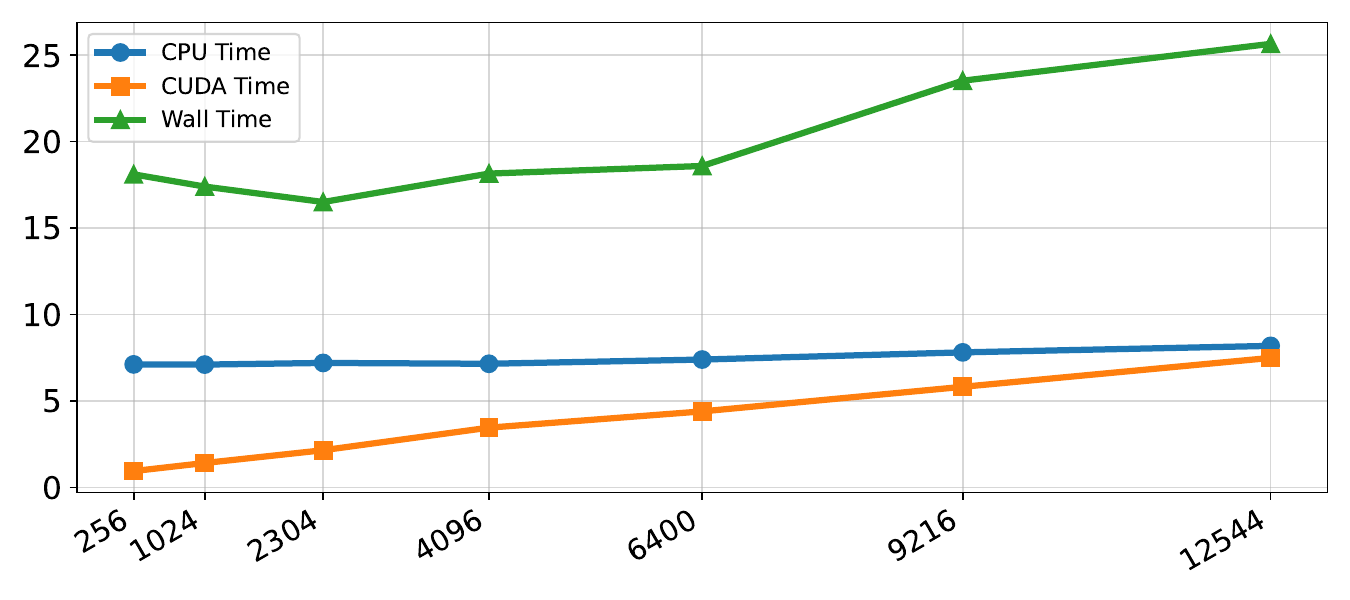}
    \end{minipage}
    \hfill
    \begin{minipage}{0.49\textwidth}
        \centering
        \includegraphics[width=\linewidth]{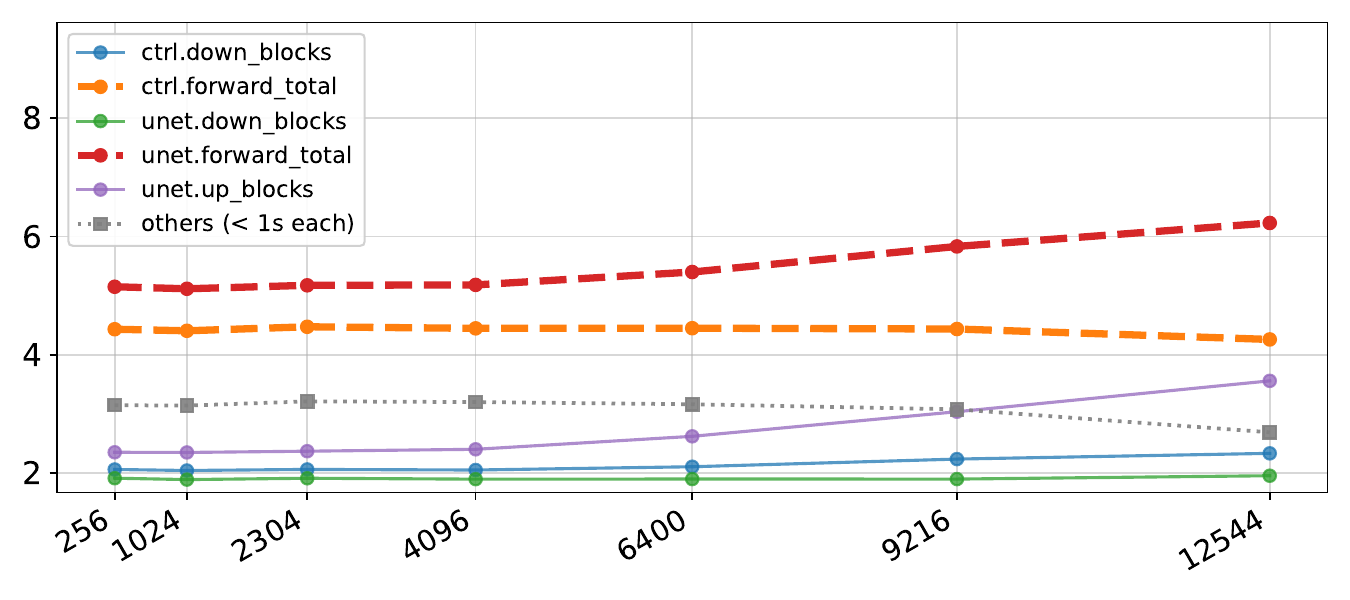}
    \end{minipage}
    \caption{
    Profiling on the RTX 6000 as a function of the output point count. Left: CPU operator, CUDA kernel and wall time. Right: per-component runtime.}
    \label{fig:profiler_times_points2}
\end{figure*}

\begin{figure*}[t]
    \centering
    \begin{minipage}{0.49\textwidth}
        \centering
        \includegraphics[width=\linewidth]{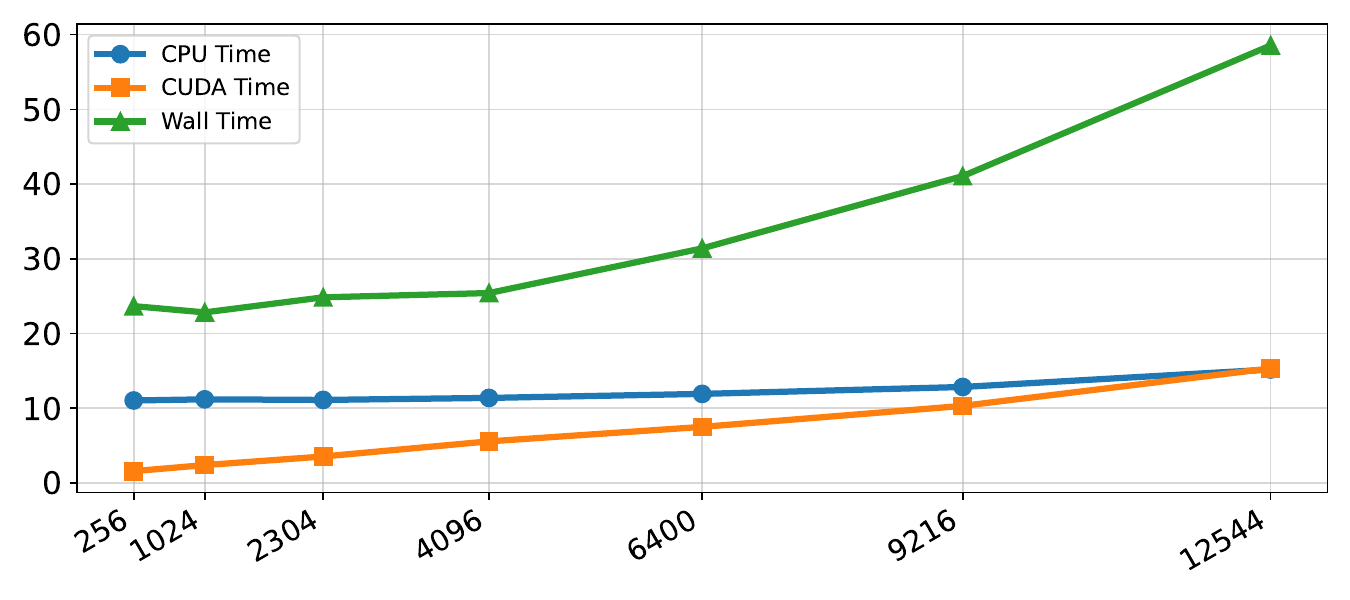}
    \end{minipage}
    \hfill
    \begin{minipage}{0.49\textwidth}
        \centering
        \includegraphics[width=\linewidth]{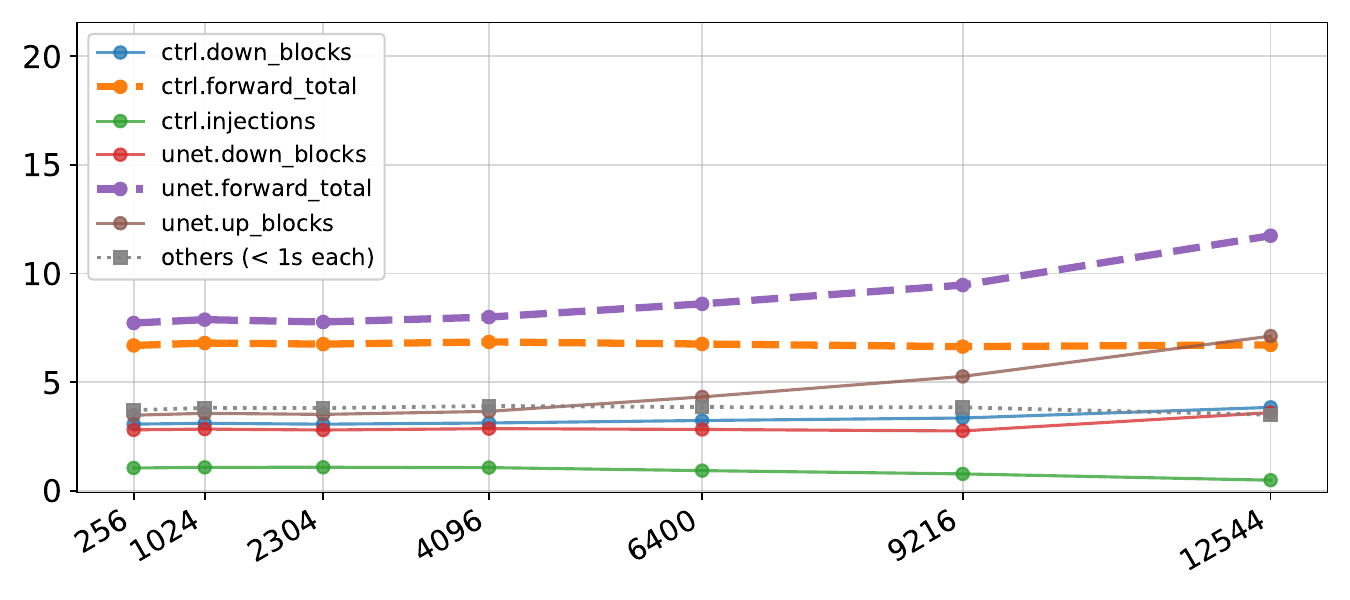}
    \end{minipage}
    \caption{
    The same profiling on a second GPU, the RTX 3090. Left: CPU operator, CUDA kernel and wall time. Right: per-component runtime.}
    \label{fig:profiler_times_points3}
\end{figure*}

The runtimes in Figure~\ref{fig:timing} were measured on one machine with an RTX 6000 GPU and an AMD EPYC 7343 CPU; WVS and BNOT run on the CPU, GBN and our method on the GPU, and each entry is the mean over 10 runs on Icons-50 targets. Here we profile a single generation while sweeping the OT grid from $16\times16$ (256 points) to $112\times112$ (12544 points).

On the CPU, denoising time grows steadily with the point count (Figure~\ref{fig:profiler_times_points1}). On both GPUs (Figures~\ref{fig:profiler_times_points2} and~\ref{fig:profiler_times_points3}, left), the CUDA kernel time grows with the grid but stays small, while the wall time is dominated by a nearly constant overhead from launching many small kernels at every denoising step. The wall time therefore stays nearly flat and rises only beyond a few thousand points, as in Figure~\ref{fig:timing}. The per-component breakdowns (right) attribute this rise to the convolutional U-Net blocks. Profiling itself inflates the absolute timings, so these values should be read for their scaling only.

\section{Extended Spectral Analysis}
\label{app:spectral}

\begin{figure*}[t]
    \centering
    \includegraphics[width=\linewidth]{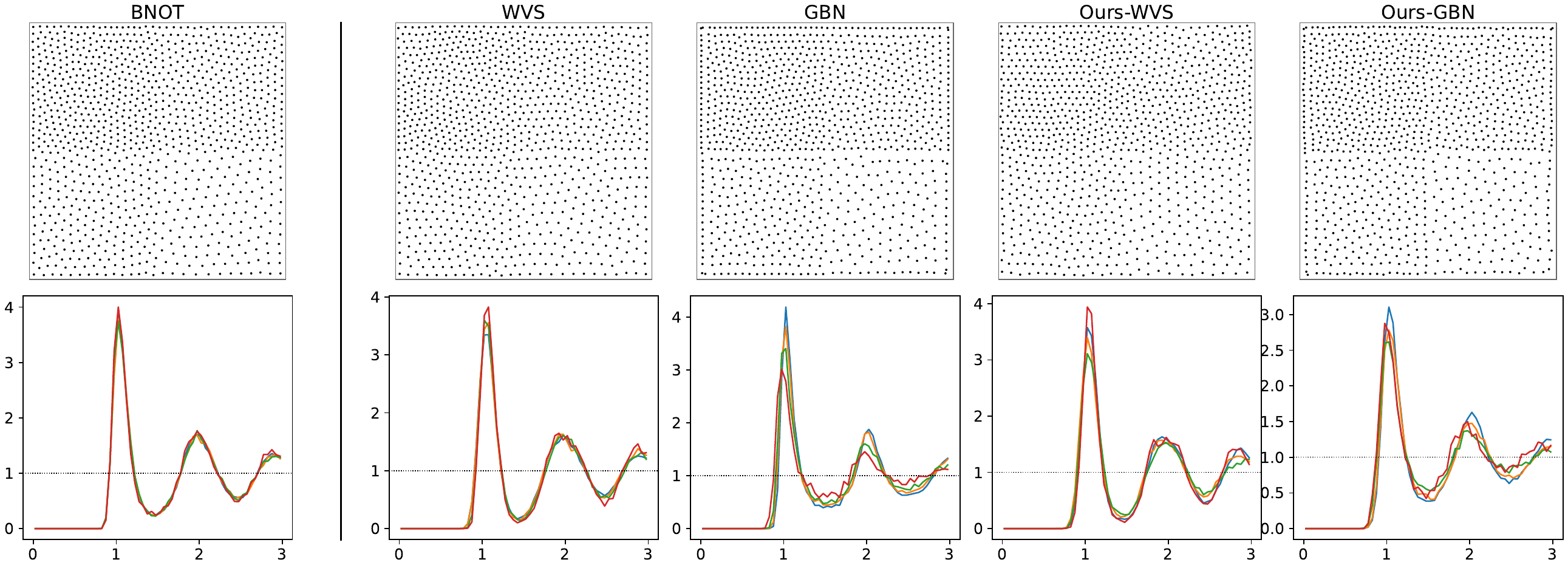}
    \caption{
    Pair correlation $g(r)$ on \textsc{quad4}, $r$ in mean spacings: $\rho = 1.00$ (blue), $0.80$ (orange), $0.60$ (green), $0.40$ (red).
    }
    \label{fig:Spectral1}
\end{figure*}

Figure~\ref{fig:Spectral1} complements the spectral analysis of Figure~\ref{fig:Spectral3} with pair correlations on the \textsc{quad4} target, with $1024$ points and $25$ realizations, computed within each constant-density region from interior centers only. Coinciding curves mean that a sampler keeps its correlation structure as the density changes. All methods keep an empty exclusion zone and a first peak at $r \approx 1$; the peak height varies least for BNOT and WVS, and less for Ours-GBN than for its exemplar GBN, but more for Ours-WVS than for WVS.

\section{Qualitative Results \& Comparisons}
\label{app:qualitative}

\begin{figure*}[t]
    \centering
    \includegraphics[width=\linewidth,height=0.92\textheight,keepaspectratio]{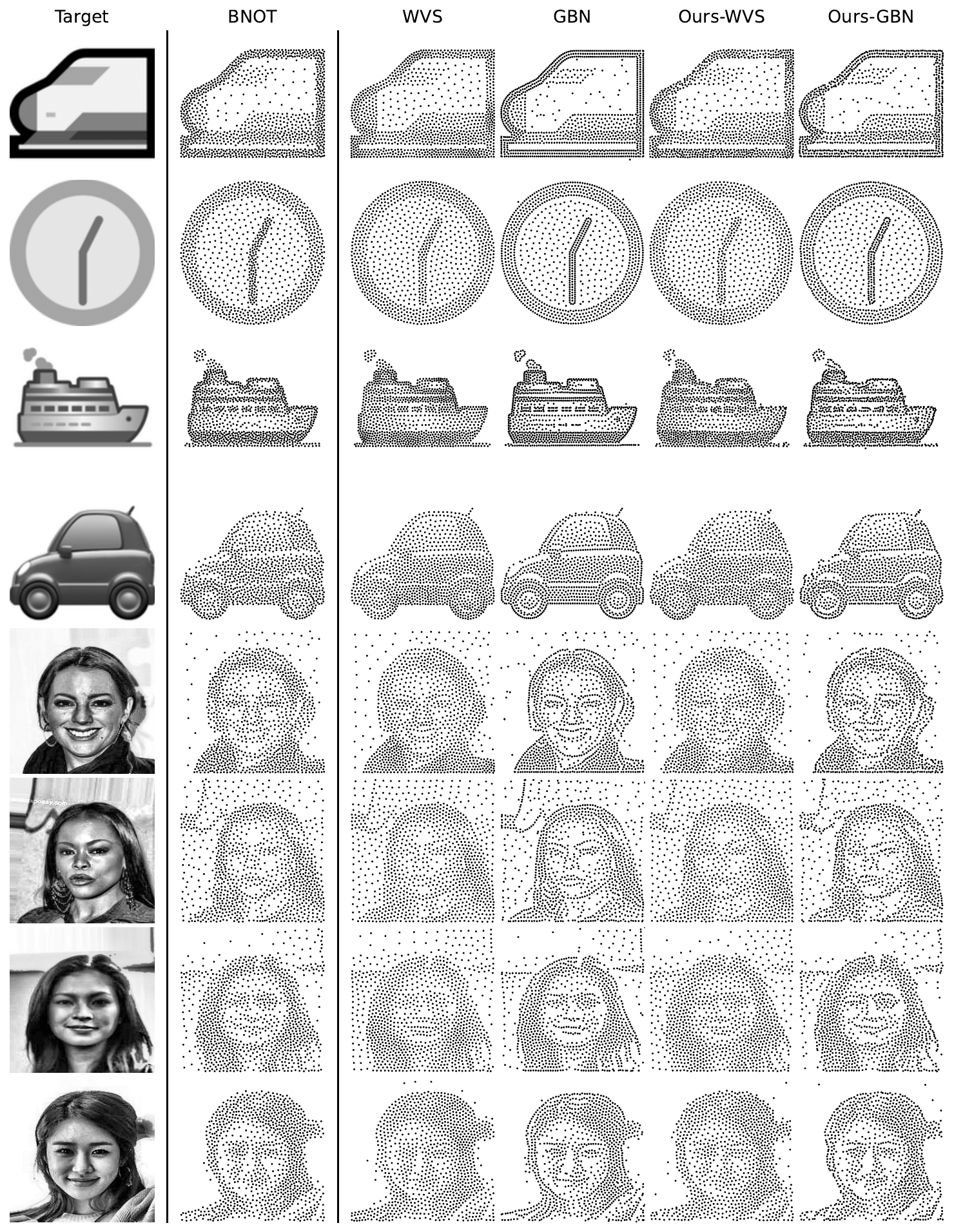}
    \caption{
    Additional comparisons with BNOT, WVS and GBN on held-out Icons-50 and CelebA images, at 1024 points.}
    \label{fig:sup_comparison1}
\end{figure*}

\begin{figure*}[t]
    \centering
    \includegraphics[width=\linewidth,height=0.92\textheight,keepaspectratio]{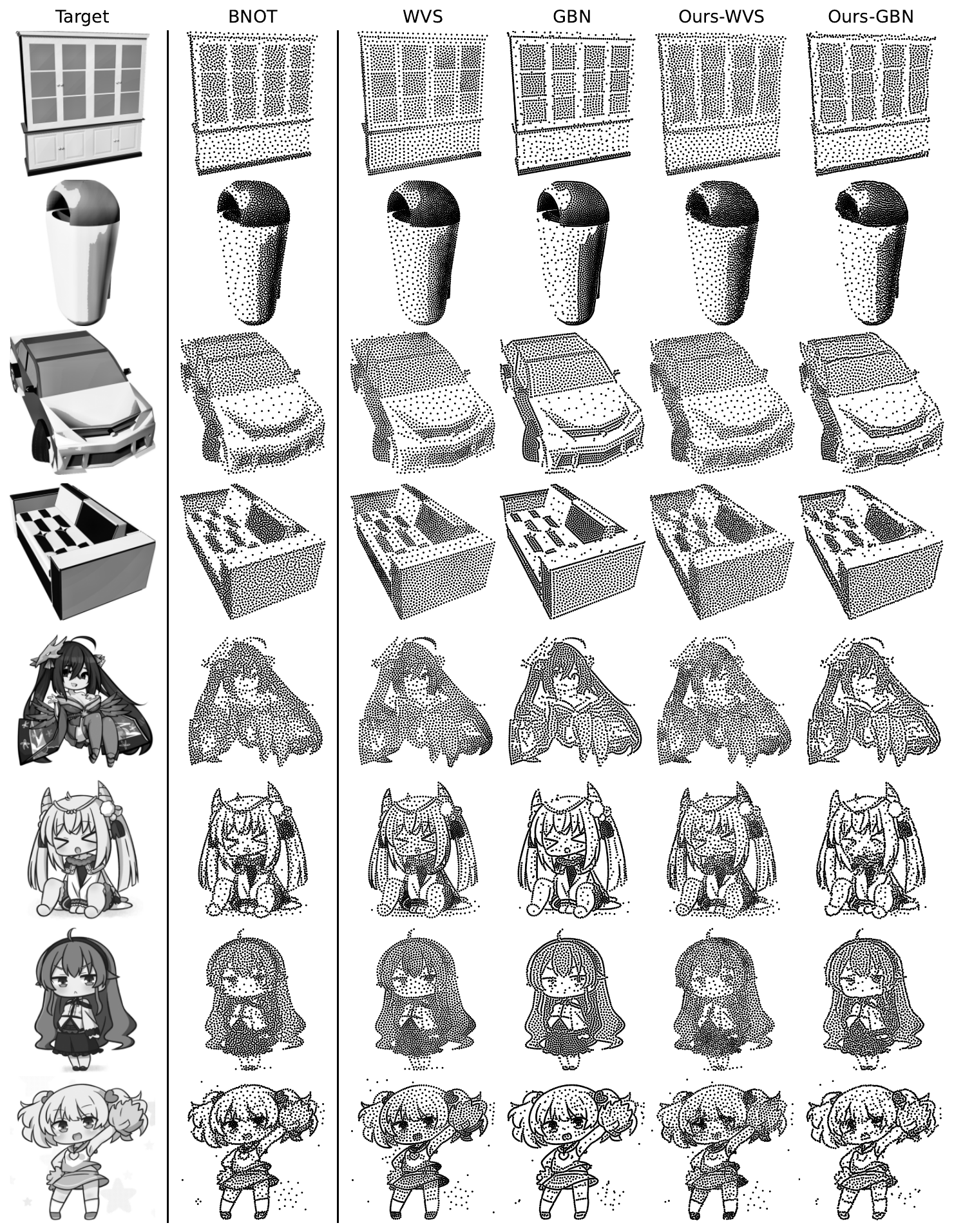}
    \caption{
    Additional comparisons with BNOT, WVS and GBN on held-out ShapeNet2D and Anime Chibi images, at 1600 points.}
    \label{fig:sup_comparison2}
\end{figure*}

Figures~\ref{fig:sup_comparison1} and~\ref{fig:sup_comparison2} show further held-out test images, complementing Figure~\ref{fig:qualitative}, each as the target image followed by BNOT, WVS, GBN and our two variants, at the point budget of its dataset.

\section{Full Quantitative Comparison}
\label{app:full_quantitative}

Table~\ref{tab:quantitative_comparison_full} repeats the comparison of Table~\ref{tab:quantitative_comparison} with the standard deviation over each test set, so that each variant can be checked against the sampler whose exemplars it was trained on, and against BNOT as an unranked reference.

\begin{table*}[b]
\centering
\small
\setlength{\tabcolsep}{4pt}
\renewcommand{\arraystretch}{1.2}
\begin{tabular}{|l|c||c|c|c|c|}
\hline
\textbf{Metric} & BNOT (ref.) & WVS & GBN & \textbf{Ours-WVS} & \textbf{Ours-GBN} \\
\hline
\multicolumn{6}{|l|}{\textbf{Icons-50 (1000 images, 1024 points)}} \\
\hline
CVT ($\times 10^{-5}$)\,$\downarrow$ & 2.355 $\pm$ 1.696 & \textbf{2.194 $\pm$ 1.583} & 2.370 $\pm$ 1.691 & \underline{2.265 $\pm$ 1.601} & 2.450 $\pm$ 1.711 \\
HOT ($\times 10^{-5}$)\,$\downarrow$ & 2.421 $\pm$ 1.734 & 10.536 $\pm$ 9.583 & \underline{5.204 $\pm$ 4.864} & 6.459 $\pm$ 4.763 & \textbf{4.059 $\pm$ 3.055} \\
Capacity $\delta_c$\,$\downarrow$ & 0.010 $\pm$ 0.004 & 0.041 $\pm$ 0.020 & \textbf{0.010 $\pm$ 0.009} & 0.041 $\pm$ 0.025 & \underline{0.018 $\pm$ 0.022} \\
Centroid $D$\,$\downarrow$ & 0.001 $\pm$ 0.000 & 0.014 $\pm$ 0.005 & \underline{0.008 $\pm$ 0.005} & 0.010 $\pm$ 0.004 & \textbf{0.006 $\pm$ 0.003} \\
EMD\,$\downarrow$ & 0.010 $\pm$ 0.002 & 0.020 $\pm$ 0.006 & \underline{0.015 $\pm$ 0.005} & 0.016 $\pm$ 0.004 & \textbf{0.013 $\pm$ 0.004} \\
Sinkhorn\,$\downarrow$ & 0.092 $\pm$ 0.003 & 0.094 $\pm$ 0.003 & \underline{0.093 $\pm$ 0.003} & \underline{0.093 $\pm$ 0.003} & \textbf{0.092 $\pm$ 0.003} \\
Spatial $\rho_s$\,$\uparrow$ & 0.526 $\pm$ 0.096 & \textbf{0.589 $\pm$ 0.110} & 0.516 $\pm$ 0.105 & \underline{0.562 $\pm$ 0.117} & 0.494 $\pm$ 0.110 \\
\hline
\multicolumn{6}{|l|}{\textbf{CelebA (500 images, 1024 points)}} \\
\hline
CVT ($\times 10^{-5}$)\,$\downarrow$ & 5.927 $\pm$ 1.678 & \textbf{5.249 $\pm$ 1.593} & 6.357 $\pm$ 1.722 & \underline{5.687 $\pm$ 1.638} & 6.766 $\pm$ 1.820 \\
HOT ($\times 10^{-5}$)\,$\downarrow$ & 6.118 $\pm$ 1.719 & 21.807 $\pm$ 9.474 & 12.864 $\pm$ 6.241 & \underline{11.030 $\pm$ 3.780} & \textbf{8.265 $\pm$ 2.493} \\
Capacity $\delta_c$\,$\downarrow$ & 0.013 $\pm$ 0.003 & 0.054 $\pm$ 0.018 & \textbf{0.013 $\pm$ 0.006} & 0.051 $\pm$ 0.022 & \underline{0.022 $\pm$ 0.011} \\
Centroid $D$\,$\downarrow$ & 0.001 $\pm$ 0.000 & 0.016 $\pm$ 0.004 & 0.010 $\pm$ 0.004 & \underline{0.009 $\pm$ 0.002} & \textbf{0.005 $\pm$ 0.002} \\
EMD\,$\downarrow$ & 0.013 $\pm$ 0.000 & 0.024 $\pm$ 0.004 & 0.018 $\pm$ 0.004 & \underline{0.017 $\pm$ 0.002} & \textbf{0.015 $\pm$ 0.002} \\
Sinkhorn\,$\downarrow$ & 0.095 $\pm$ 0.001 & 0.097 $\pm$ 0.001 & \underline{0.096 $\pm$ 0.001} & \underline{0.096 $\pm$ 0.001} & \textbf{0.095 $\pm$ 0.001} \\
Spatial $\rho_s$\,$\uparrow$ & 0.659 $\pm$ 0.070 & \textbf{0.741 $\pm$ 0.073} & 0.604 $\pm$ 0.069 & \underline{0.718 $\pm$ 0.082} & 0.599 $\pm$ 0.074 \\
\hline
\multicolumn{6}{|l|}{\textbf{ShapeNet2D (400 images, 1600 points)}} \\
\hline
CVT ($\times 10^{-5}$)\,$\downarrow$ & 3.070 $\pm$ 1.511 & \textbf{2.629 $\pm$ 1.437} & \underline{3.391 $\pm$ 1.562} & 3.530 $\pm$ 1.824 & 3.688 $\pm$ 1.667 \\
HOT ($\times 10^{-5}$)\,$\downarrow$ & 3.206 $\pm$ 1.547 & 11.432 $\pm$ 5.604 & \underline{7.055 $\pm$ 3.918} & 7.253 $\pm$ 2.822 & \textbf{5.367 $\pm$ 1.915} \\
Capacity $\delta_c$\,$\downarrow$ & 0.013 $\pm$ 0.006 & 0.043 $\pm$ 0.023 & \textbf{0.013 $\pm$ 0.009} & 0.059 $\pm$ 0.057 & \underline{0.038 $\pm$ 0.045} \\
Centroid $D$\,$\downarrow$ & 0.001 $\pm$ 0.000 & 0.012 $\pm$ 0.003 & \underline{0.007 $\pm$ 0.004} & 0.008 $\pm$ 0.003 & \textbf{0.006 $\pm$ 0.003} \\
EMD\,$\downarrow$ & 0.010 $\pm$ 0.001 & 0.019 $\pm$ 0.005 & \underline{0.015 $\pm$ 0.004} & \underline{0.015 $\pm$ 0.004} & \textbf{0.013 $\pm$ 0.003} \\
Sinkhorn\,$\downarrow$ & 0.094 $\pm$ 0.002 & \underline{0.095 $\pm$ 0.002} & \textbf{0.094 $\pm$ 0.002} & \underline{0.095 $\pm$ 0.002} & \textbf{0.094 $\pm$ 0.002} \\
Spatial $\rho_s$\,$\uparrow$ & 0.539 $\pm$ 0.091 & \textbf{0.600 $\pm$ 0.103} & 0.534 $\pm$ 0.106 & \underline{0.574 $\pm$ 0.106} & 0.521 $\pm$ 0.108 \\
\hline
\multicolumn{6}{|l|}{\textbf{Anime Chibi (263 images, 1600 points)}} \\
\hline
CVT ($\times 10^{-5}$)\,$\downarrow$ & 0.847 $\pm$ 0.662 & \textbf{0.738 $\pm$ 0.581} & \underline{0.913 $\pm$ 0.697} & 1.146 $\pm$ 1.233 & 1.307 $\pm$ 1.344 \\
HOT ($\times 10^{-5}$)\,$\downarrow$ & 1.300 $\pm$ 1.491 & 4.655 $\pm$ 3.642 & 3.262 $\pm$ 2.499 & \underline{3.089 $\pm$ 2.029} & \textbf{2.608 $\pm$ 1.786} \\
Capacity $\delta_c$\,$\downarrow$ & 0.040 $\pm$ 0.104 & \underline{0.061 $\pm$ 0.024} & \textbf{0.030 $\pm$ 0.023} & 0.129 $\pm$ 0.081 & 0.107 $\pm$ 0.076 \\
Centroid $D$\,$\downarrow$ & 0.002 $\pm$ 0.004 & 0.011 $\pm$ 0.004 & 0.010 $\pm$ 0.005 & \underline{0.009 $\pm$ 0.003} & \textbf{0.007 $\pm$ 0.003} \\
EMD\,$\downarrow$ & 0.009 $\pm$ 0.005 & 0.017 $\pm$ 0.005 & 0.016 $\pm$ 0.005 & \underline{0.014 $\pm$ 0.004} & \textbf{0.013 $\pm$ 0.004} \\
Sinkhorn\,$\downarrow$ & 0.093 $\pm$ 0.002 & \underline{0.094 $\pm$ 0.002} & \underline{0.094 $\pm$ 0.002} & \textbf{0.093 $\pm$ 0.002} & \textbf{0.093 $\pm$ 0.002} \\
Spatial $\rho_s$\,$\uparrow$ & 0.422 $\pm$ 0.085 & \textbf{0.469 $\pm$ 0.096} & 0.391 $\pm$ 0.081 & \underline{0.433 $\pm$ 0.099} & 0.368 $\pm$ 0.088 \\
\hline
\end{tabular}
\caption{Quantitative comparison of WVS, GBN, and our two variants, Ours-WVS and Ours-GBN, trained on WVS and GBN exemplars respectively, across four benchmarks. Mean $\pm$ standard deviation over the respective test sets at each benchmark's point budget; best and second-best means per row are shown in bold and underlined, respectively. Each benchmark uses its corresponding trained checkpoints. BNOT is included as a reference and is not ranked, as our variants were not trained on BNOT exemplars due to its performance.
}
\label{tab:quantitative_comparison_full}
\end{table*}

\end{document}